\documentclass{article}

\usepackage{arxiv}

\usepackage[utf8]{inputenc} 
\usepackage[T1]{fontenc}    
\usepackage{hyperref}       
\usepackage{url}            
\usepackage{booktabs}       
\usepackage{amsfonts}       
\usepackage{nicefrac}       
\usepackage{microtype}      
\usepackage{lipsum}
\usepackage{graphicx}

\title{Effect of right censoring bias on survival analysis}

\author{
    Enrique Barrajón
   \\
    Department of Medical Oncology \\
    Hospital General Universitario de Elche \\
  Spain \\
  \texttt{\href{mailto:enrique.barrajon@gmail.com}{\nolinkurl{enrique.barrajon@gmail.com}}} \\
   \And
    Laura Barrajón
   \\
    Department of Hospital Pharmacy \\
    Hospital Universitario San Juan de Alicante \\
  Spain \\
  \texttt{} \\
  }

\begin{document}
\maketitle

\def\tightlist{}

\begin{abstract}
Kaplan-Meier survival analysis represents the most objective measure of
treatment efficacy in oncology, though subjected to potential bias,
which is worrisome in an era of precision medicine. Independent of the
bias inherent to the design and execution of clinical trials, bias may
be the result of patient censoring, or incomplete observation. Unlike
disease/progression free survival, overall survival is based on a well
defined time point and thus avoids interval censoring, but
right-censoring, due to incomplete follow-up, may still be a source of
bias. We study three mechanisms of right-censoring and find that one of
them, surrogate of patient lost to follow-up, is able to impact
Kaplan-Meier survival, improving significantly the estimation of
survival in comparison with complete follow-up datasets, as measured by
the hazard ratio. We also present two bias indexes able to signal
datasets with right-censoring associated overestimation of survival.
These bias indexes can detect bias in public available datasets.
\end{abstract}


\hypertarget{introduction}{%
\section{Introduction}\label{introduction}}

More than one hundred years after the statement that certainty in
science requires the confirmation of insights by measurement
\cite{Laesecke2002}, the scientific community is acknowledging a
reproducibility crisis in the last few years \cite{Baker2016}.

Clinical trials represent an area of biomedical research affected by the
replication crisis. About one third of randomized clinical trials
subjected to reanalyses have shown different conclusions to the original
article \cite{Ebrahim2014}, and around 10\% of trial conclusions
reported to the Food and Drug Administration (FDA) show some kind of
discordance to the published articles \cite{Rising2008}, and, for this
reason, it has been claimed that most of research resources are wasted
for this reason \cite{Ioannidis2014}. It is worrisome that, in an age of
precision medicine, clinical research is threatened by different kinds
of biases inherent to the design, implementation, and analysis of
clinical trials, compromising the translation of trial outcomes into
clinical benefit \cite{Heneghan2017}. It has been postulated that even a
totally unbiased, perfectly designed and executed clinical trial may
still produce false and irreproducible results \cite{Hanin2017}.

Survival analysis, as described by Kaplan and Meier, extract the most
information of incomplete observations (censoring) and represents the
most objective measure of treatment efficacy in oncology, medical
sciences in general, and engineering \cite{KaplanMeier1958}. Censoring
means the survival time to event for a subject cannot be accurately
determined because dropout, lost to follow-up, unavailable information,
or the study ends before the event of interest occurs, and may affect
the estimation of survival rate \cite{Rich2010}.

To estimate survival as a function of time, Kaplan-Meier analysis
arranges all the \(n\) study cases, sorted by the time of follow-up,
giving \(n\) intervals. Starting with \(S_{time = 0} = 1\), the
probability of survival is recalculated at every event, with the product
of the probability of survival at the previous interval and the survival
ratio in the interval: surviving patients at the end of the interval
divided by the number of patients that entered this interval (the
probability of survival is not updated at censoring times.)

\[
\large S_i = \prod_{i=1}^{n} \Big(1 - \frac{events_i}{atRisk_i} \Big)
\]

While almost half of Kaplan-Meier analyses published in medical journals
are susceptible to \emph{competing risk bias} that may overestimate
event risk \cite{vanWalraven2016}, \emph{informative censoring bias} may
be associated with overestimation of treatment effects
\cite{McNamee2017}. Censoring may change the conclusions of clinical
trials and may challenge the validity of results from interim analyses
in clinical trials \cite{Prased2015, Bagust2018}.

One of the problems with survival analysis is the accuracy of the start
and event times, which may give rise to interval censoring
\cite{Zhang2020}. This uncertainty is reduced in clinical trials with a
perfect definition of the patient entry and event dates, but, sometimes,
the event time only can be assigned within an interval, as when we
measure disease free survival or progression free survival. Overall
survival, otherwise, is based on a well defined time point and, thus,
avoids interval censoring, but it may still be the subject of
right-censoring when we do not know the event time point. Another
drawback of estimating overall survival is that more patients and time
are needed, and others factors, or further treatments, may affect
outcomes \cite{Leung1997, Saad2016, Sridhara2013, Nunan2018}.

The Kaplan Meier survival curve, as well as the logrank test, are based
on the assumptions that censoring is unrelated to prognosis (the random
censoring model), the survival probability is independent of the
chronological time of inclusion in the study, and events happen at the
times specified. Deviations from these assumptions may affect results if
they are satisfied differently in the groups being compared, for example
if censoring is more likely in one group than another \cite{Bland2004}.
It has been stated that elimination from the database of patients lost
to follow-up or who withdraw consent may affect the power and validity
of conclusions derived from the clinical study \cite{Gabriel2011}.
Validity of the random censorship model, that we take for granted, may
not be proved nor excluded with a statistical test, and might lead to
biased results of the Kaplan-Meier estimator \cite{Krstajic2017}.

In this project, the \emph{R language} \cite{R2018}, the integrated
development environment \emph{RStudio} \cite{RStudio2020}, the
\emph{survival} \cite{survival2015}, and \emph{OptimalCutpoints}
\cite{Lopez2014} packages were used to simulate population time to
event, censoring time, and to the systematic study of bias, as well as
the potential bias in public available datasets \cite{survivalDatasets}.

\hypertarget{virtual-datasetes}{%
\section{Virtual datasetes}\label{virtual-datasetes}}

The \emph{survival} package uses datasets with, at least, a vector of
times, and a vector of states, the latter informing the state of every
case: event or censored. To study the effect of censoring on survival,
we simulate datasets with complete follow-up and derive from them
datasets with different patterns of censoring.

\hypertarget{complete-follow-up}{%
\subsection{Complete follow-up}\label{complete-follow-up}}

The survival function in patients with cancer, or any other condition,
may be simulated according to a Weibull mixture cure-rate model, where
the survival probability is the result of a subpopulation of cases with
events (deaths, relapses, etc.) occurring following a Weibull
distribution, plus a subpopulation of cured cases, never experiencing
the event \cite{Othus2012}. The cured fraction has been defined as the
proportion of patients with the same death rates of the general
population. A problem with this definition is that the time when the
death rate is similar to the general population differs by tumor type,
and is much longer than the usual duration of studies, and the normal
approach is to talk about the proportion of long-term survival (at some
arbitrary time interval) \cite{DalMaso2014}.

A hypothetical dataset with complete follow-up may be approached by a
general 2-parameter Weibull reliability/survivorship function with shape
parameter \(k \in (0, \infty)\) and scale parameter
\(b \in (0, \infty)\) \cite{Siegrist2018}. The exponential function is a
particular case of the Weibull function with \(k = 1\).

\[
\large S(t) = e^{-\big(\frac{t}{b}\big)^k}
\]

The inverse Weibull distribution allows to simulate a time distribution
\(t\) knowing the parameters \(b\) and \(k\), and a random probability
of survival \(p \in (0, 1)\).

\[
\large t = b \; \big(ln \frac{1}{p}\big)^{1/k}
\]

Parameters \(k\) and \(b\) have no practical meaning, and a more
convenient way of representing the 2-parameter Weibull survival function
is as a function of quantiles, which can be measured directly in
survival datasets. The quantiles in the Weibull cumulative distribution
are defined by the expression:

\[
  \large Q_p = b \; [-ln(1-p)]^{1/k}
\]

We may use this expression with two known quantiles, such as the
\emph{median}, \(Q_{50}\), and the 99th percentile, \(Q_{99}\),

\[
  \Large Q_{50} = b \; (ln \; 2)^{1/k}
\]

\[
  \Large Q_{99} = b \; (ln \; 100)^{1/k}
\]

to obtain the parameters shape \(k\) and scale \(b\) of a Weibull
distribution:

\[
k = \frac{ln(ln \; 100 \; - \; ln \; 2)} {ln \; Q_{99} \; - \; ln \; Q_{50}}
\]

\[
  \large b = \; \frac{Q_{50}}{(ln \;2)^{1/k}}
\]

Using these tools, we may obtain a dataset with vector of times and
states. The vector of times is obtained from a vector of random
probabilities and the parameters of a Weibull mixture cure-rate
distribution: the median event time, the 99th percentile event time, and
a given cure-rate for the population. For convenience, the 99th
percentile event time in the virtual datasets is fixed to 100 units of
time and we let the median survival vary in the range 5-95 units of
time; the cure rate is simulated in the arbitrary range 0-80\%. For a
complete follow-up dataset, the states vector is 0 for the cured cases,
and 1 all the other cases, event cases.

\hypertarget{incomplete-follow-up}{%
\subsection{Incomplete follow-up}\label{incomplete-follow-up}}

We may obtain censored datasets from a complete follow-up dataset by one
of three different mechanisms of censoring:

\renewcommand{\theenumi}{\Alph{enumi}}
\begin{enumerate}

\item \textbf{Time censoring}

For every case in the complete follow-up dataset, a new \textit{random time} is generated from a Weibull distribution with $median$ and $Q_{99}$ percentile times, and a proportion of censoring. Every case is updated if this new time is shorter than the complete follow-up time: the new time replaces the complete follow-up one and the state is updated to 0 (censored).

\item \textbf{Interim censoring}

A random \textit{interim time} is obtained from a Weibull distribution with $median$ and $Q_{99}$ percentile times. For every case in the complete follow-up dataset, a random \textit{recruit time} is generated. If the interval between \textit{recruit time} and \textit{interim time} is shorter than the complete follow-up time, this interval replaces the complete follow-up time and the state is changed to 0 (censored). This mechanism parallels the censoring associated with interim analysis.

\item \textbf{Case censoring}

A random sample of cases is selected from the complete follow-up dataset with a given probability of censoring. For these censored cases, the corresponding time is shortened a random amount, and the corresponding state is changed to 0 (censored). This mechanism of censoring assimilates retired cases from clinical trials, either lost to follow-up or who withdraw consent.

\end{enumerate}

\hypertarget{virtual-trials}{%
\section{Virtual trials}\label{virtual-trials}}

A virtual trial is constituted by a censored dataset and the
corresponding complete follow-up dataset. The complete follow-up dataset
is simulated with one thousand cases following a Weibull mixture
cure-rate model. The censored dataset is obtained from the complete
follow-up dataset, applying one of the three described mechanisms of
censoring. The virtual trial allows to study the impact of censoring on
survival, by estimating the hazard ratio (\emph{HR}) and \emph{p-value}
of the censored dataset compared with the complete follow-up dataset. A
hazard ratio of 1 means that the hazard of the censored dataset and the
complete follow-up dataset are the same, while a hazard ratio less than
1 means the hazard of the censored dataset is less than the hazard of
the complete follow-up dataset; a p-value less than 0.05 means a low
probability that the results are due to chance \cite{Steyerberg2010}.
The complete follow-up and censored Kaplan-Meier survival curves may be
represented together.

\begin{figure}[ht]
  \centering
    \includegraphics[width=1\textwidth]{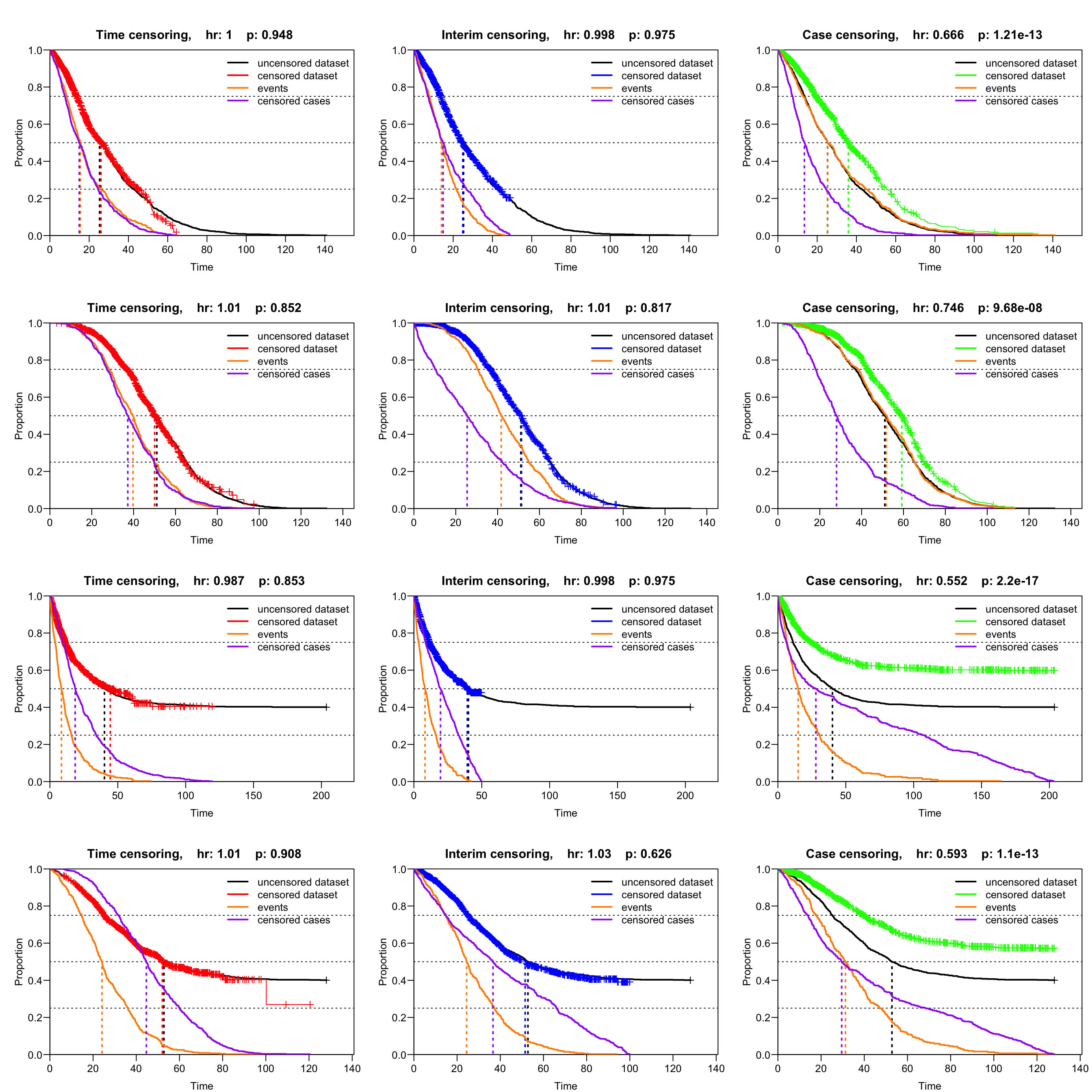}
  \caption{Influence of censoring type on Kaplan-Meier estimate (see text)}
  \label{fig1}
\end{figure}

Figure \ref{fig1} shows 12 virtual experiments based on 4 complete
follow-up datasets with median event time of 25 and 50 time units
(\(Q_{99}\) percentil of 100 time units), and cure-rate of 0 and 0.4.
For every complete follow-up dataset, three censored datasets are
obtained with the three different mechanisms of censoring and proportion
of censoring of 50\%. For every virtual trial, we plot the complete
follow-up dataset in black, and the censored dataset in color: red for
\emph{time censoring}, blue for \emph{interim censoring}, and green for
\emph{case censoring}. In addition, for the censored dataset, we
represent differentially the survival curves of event (\(\mathcal{E}\))
and censored (\(\mathcal{C}\)) cases in the events interval
\(t \in (0, max(\mathcal{E}))\): event times in orange, and censored
times in purple, to find out patterns that may be associated with a
potential change in survival.

While time censoring and interim censoring survival curves overlap with
the complete follow-up survival curve, with HR around 1, case censoring
shows an increase in survival signaled by HR below 1. An interactive
simulation is presented at
\url{https://adio.shinyapps.io/right_censoring}.

\hypertarget{virtual-experiments}{%
\section{Virtual experiments}\label{virtual-experiments}}

\begin{figure}[ht]
  \centering
    \includegraphics[width=0.6\textwidth]{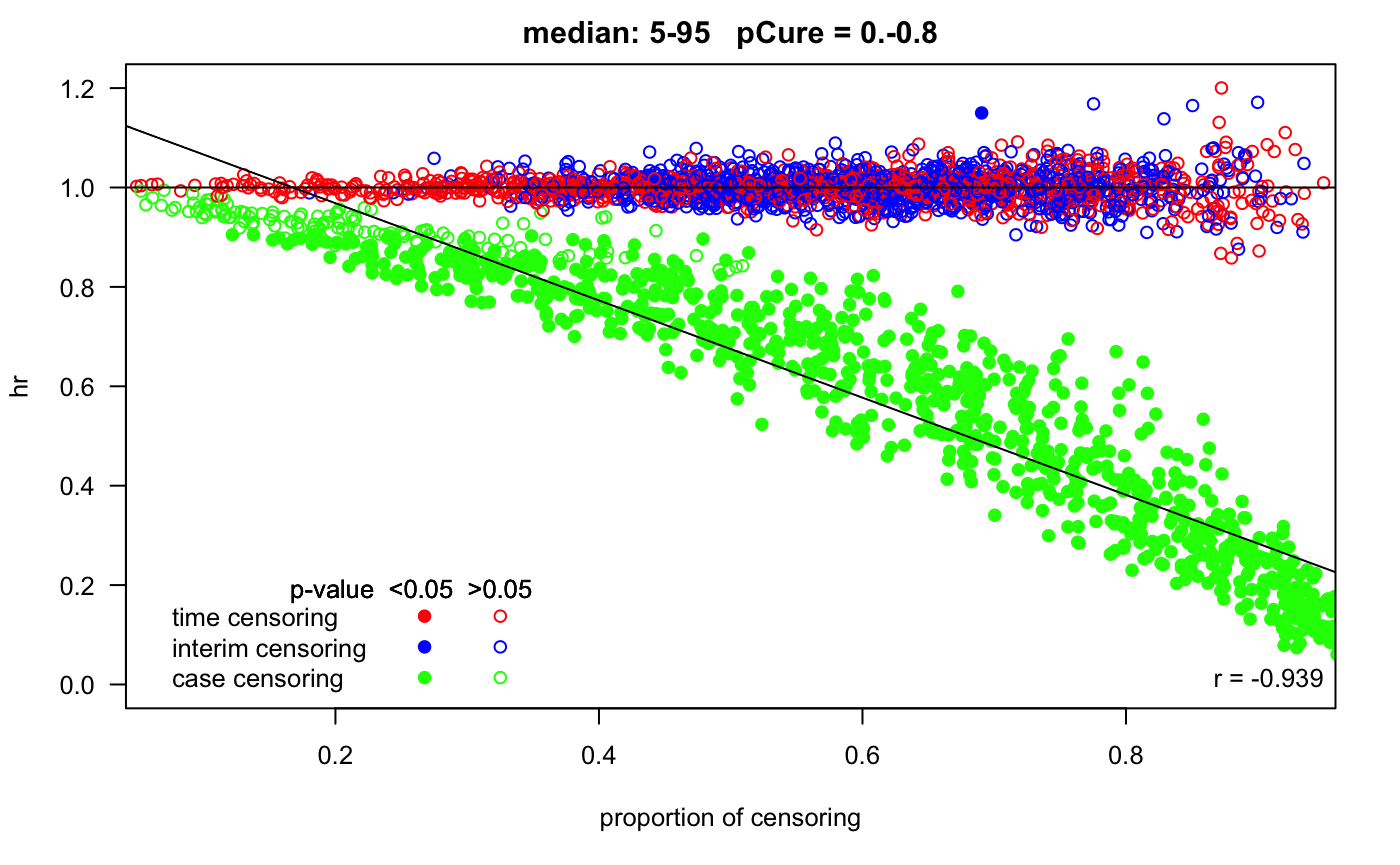}
  \caption{Influence of censoring proportion on survival estimation}
  \label{fig2}
\end{figure}

Virtual experiments are composed of hundreds of virtual trials. In a
typical virtual experiment, 1000 random complete follow-up datasets of
1000 cases are simulated with a fixed \(Q_{99}\) percentil of event
times of 100 units and random values of median event time in the range 5
to 95 units, and random cure-rate in the range 0-0.8. For every complete
follow-up dataset, three censored datasets are simulated with each of
the described mechanisms of censoring and random proportion of censoring
in the range of 5\% to 95\%, resulting in 3000 virtual trials. Every
virtual trial may be represented as a dot in a scatter plot: time
censoring (red), interim censoring (blue) and case censoring (green).
Figure \ref{fig2} shows a scatter plot of HR vs the proportion of
censoring; solid dots represent trials with a significant change in HR
(p-value \textless{} 0.05). We may observe how, unlike time censoring
and interim censoring, case censoring is associated with a biased
estimation of survival: HR correlates inversely with the proportion of
censoring (r = -0.939).

\hypertarget{bias-measures}{%
\subsection{Bias measures}\label{bias-measures}}

\begin{figure}[ht]
  \centering
    \includegraphics[width=1\textwidth]{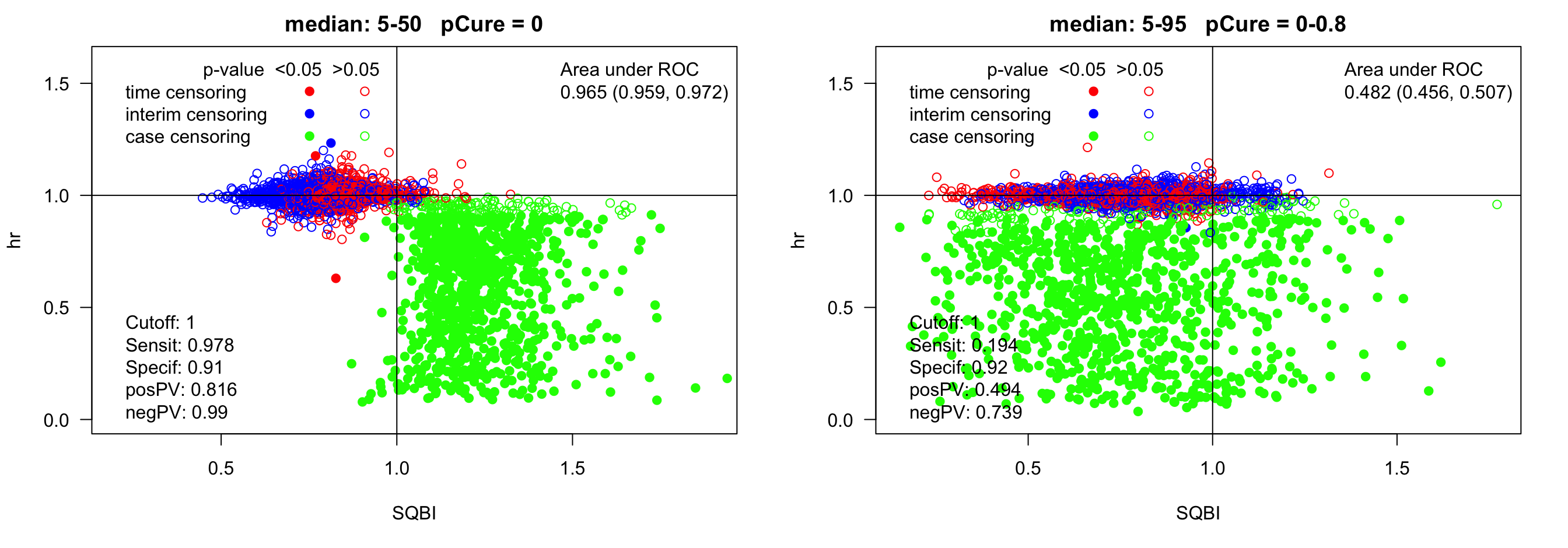}
  \caption{Influence of SQBI on survival estimation for right skewed distributions with cure-rate = 0 (left panel) or overall (right panel)}
  \label{fig3}
\end{figure}

To study bias, virtual experiments were simulated with thousands of
virtual trials with random Weibull mixture cure-rate models. For every
virtual trial, we estimate the HR and p-value of the censored to
complete follow-up datasets, and a \emph{bias index} (BI). A pragmatic
approach was used to find a BI able to signal virtual trials decreasing
the HR with a statistically significant p-value. Scatter plots of HR vs
BI represent every virtual trial as a dot (trials based on time
censoring in red, interim censoring in blue, and case censoring in
green; solid circles denote p-value \textless{} 0.05) {[}Fig.
\ref{fig3}{]}.

In order to obtain a cutoff for the BI to maximize the sensitivity and
specificity to classify virtual trials as biased (p-value \textless{}
0.5) or unbiased, the Receiver Operating Characteristic (ROC) and the
Youden index were estimated with the \emph{OptimalCutpoints} package
\textbackslash cite\{Lopez2014{]}. The obtained cutoff is then used to
rescale the BI to obtain a scaled BI with cutoff of 1, a convenient
threshold indicating that the virtual trial may be biased.

\hypertarget{quantile-bias-index-qbi}{%
\subsubsection{Quantile bias index
(QBI)}\label{quantile-bias-index-qbi}}

Biological distributions, and survival distributions in particular, are
frequently right-skewed, as common measures, such as time, have a lower
bound of zero \cite{NIST2012}. A first glance at right skewed virtual
trials of figure \ref{fig1} shows that the observed significant decrease
in HR found in case censoring trials (green curves) might be associated
with the separation of events and censored subgroups of the censored
dataset in the right-most part of their survival curves. A QBI function
was constructed based on the ratio of 95th quantiles of event times and
censor times:

\[
  \Large {QBI} = \frac{ Q_{0.95}({\mathcal{E}}) } {Q_{0.95}(\mathcal{C})}
\]

Figure \ref{fig3} shows 2 experiments, each of them with 3000 trials
based on 1000 complete follow-up datasets and the 3 mechanisms of
censoring described, with random proportion of censoring in the range
0.05 to 0.95. QBI differentiates pretty well for right skewed
distributions (datasets with median event time below 50 unit times) and
cure-rate of 0 (figure \ref{fig3} left panel). A QBI cutoff of 1.2
separates efficiently virtual trials with a significant decrease in HR
from virtual trials with HR around 1, with an area under the Receiver
Operating Characteristic (ROC) curve of 0.965, and 95\% confidence
interval (CI) of 0.959-0.972. A scaled QBI (SQBI) is calculated dividing
the QBI by the cutoff found:

\[
  \Large {SQBI} = \frac{ QBI } {{cutoff}}
\]

SQBI, nevertheless, fails with trials based on symmetric or left skewed
distributions of events, or when there are a proportion of long-term
survival cases. The estimated area under the ROC curve is 0.49 (95\% CI:
0.465-0.515) for trials with median event times in the range of 5-95
units and cure-rate of 0-80\% (figure \ref{fig3} right panel),
equivalent to diagnose bias by pure chance.

\hypertarget{under-the-mean-bias-index-umbi}{%
\subsubsection{Under the mean bias index
(UMBI)}\label{under-the-mean-bias-index-umbi}}

A more consistent BI is based on the proportion of censored cases in the
event interval, \(t \in (0, max(\mathcal{E}))\), which are under the
mean event time.

\[
  \Large {UMBI} = \frac{|\mathcal{C}\;  <\; \overline{\mathcal{E}}|} {|\mathcal{C}\;  <\; \max({\mathcal{E}})|}
\]

\begin{figure}[ht]
  \centering
  \includegraphics[width=1\textwidth]{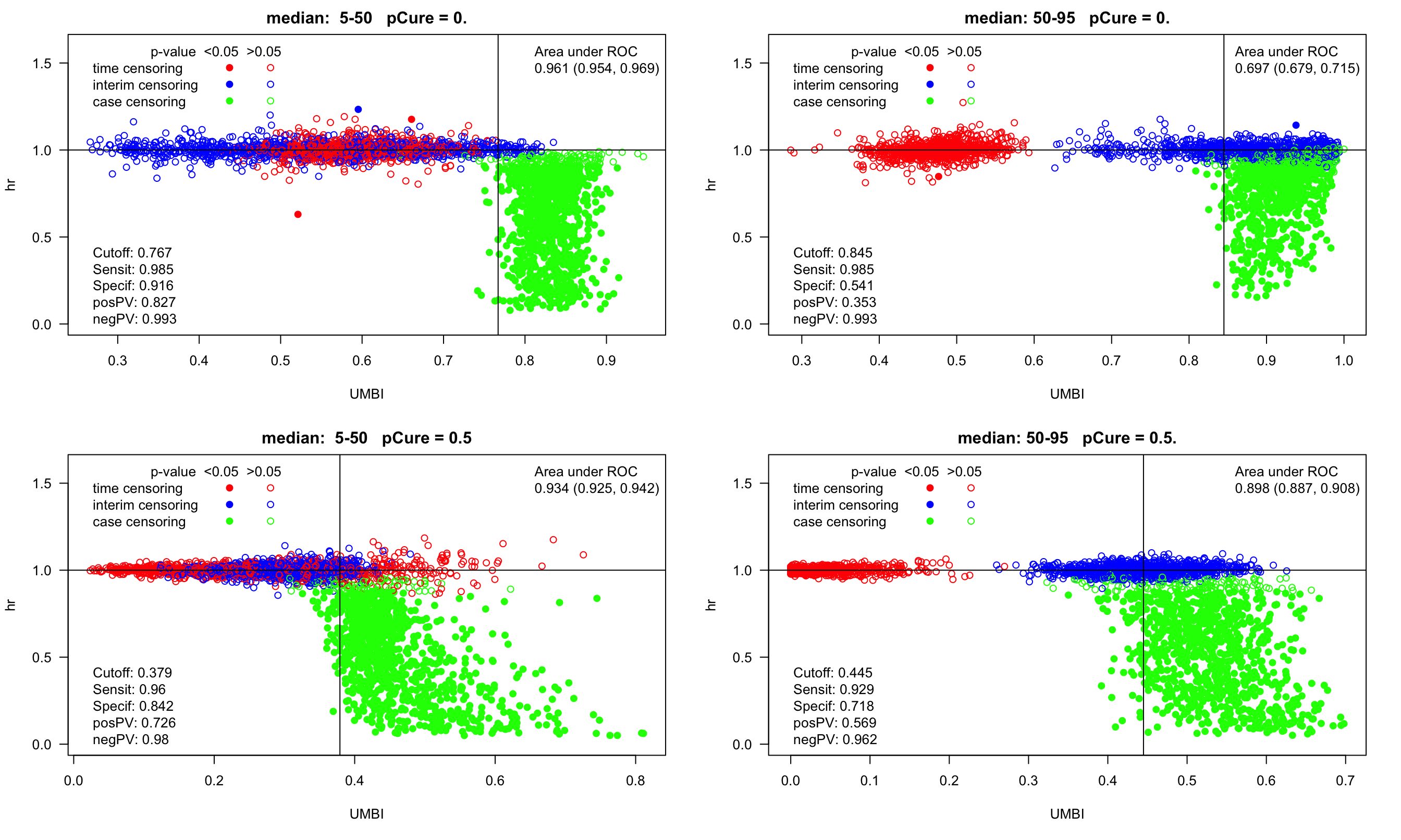}
  \caption{Influence of UMBI on survival estimation}
  \label{fig4}
\end{figure}

Figure \ref{fig4} represents 4 virtual experiments with a 2x2 design:
right skewed Weibull distributions or not, and cure-rate of 0 or 50\%.
UMBI separates virtual trials with a significant decrease in HR from
virtual trials with HR around 1, with area under the ROC curve over
0.703 (0.961 for right skewed distributions and no cure-rate), but the
cutoff is not consistent: it increases with the median event time and
decreases with the proportion of long term survivors. Thus, we need to
normalize this index to get a consistent index independent of median
event time and cure-rate. Since the cure-rate may be unreliable to be
estimated from an incomplete follow-up dataset, the long-term proportion
of cases after the last event is used.

\newpage

\hypertarget{adjusted-bias-index-abi}{%
\subsubsection{Adjusted bias index
(ABI)}\label{adjusted-bias-index-abi}}

To normalize UMBI, we need to adjust for survival skewness and the
proportion of long-term survival cases, multiplying by two factors to
obtain the pursued ABI:

\begin{itemize}

\item Multiplying by the ratio of mean $\overline{\mathcal{C}}$ to median $\widetilde{{\mathcal{C}\;}}$ censored times, corrects for the effect of skewness in the distribution.

\item Multiplying by $e$ to the power of the estimator of Kaplan-Meier long-term survival $\widehat{S}_{KM(t = \max(\mathcal{E}))}$ corrects for the effect of the long-term survival rate.

\end{itemize}

\[
  \Large ABI = \frac{|\mathcal{C}\;  <\; \overline{\mathcal{E}}|} {|\mathcal{C}\;  <\; \max({\mathcal{E}})|} \; \frac{\overline{\mathcal{C}}}{\widetilde{{\mathcal{C}\;}}} \; \LARGE{e}^{\widehat{S}_{KM(t = \max(\mathcal{E}))S}}
\]

\begin{figure}[ht]
  \centering
    \includegraphics[width=1\textwidth]{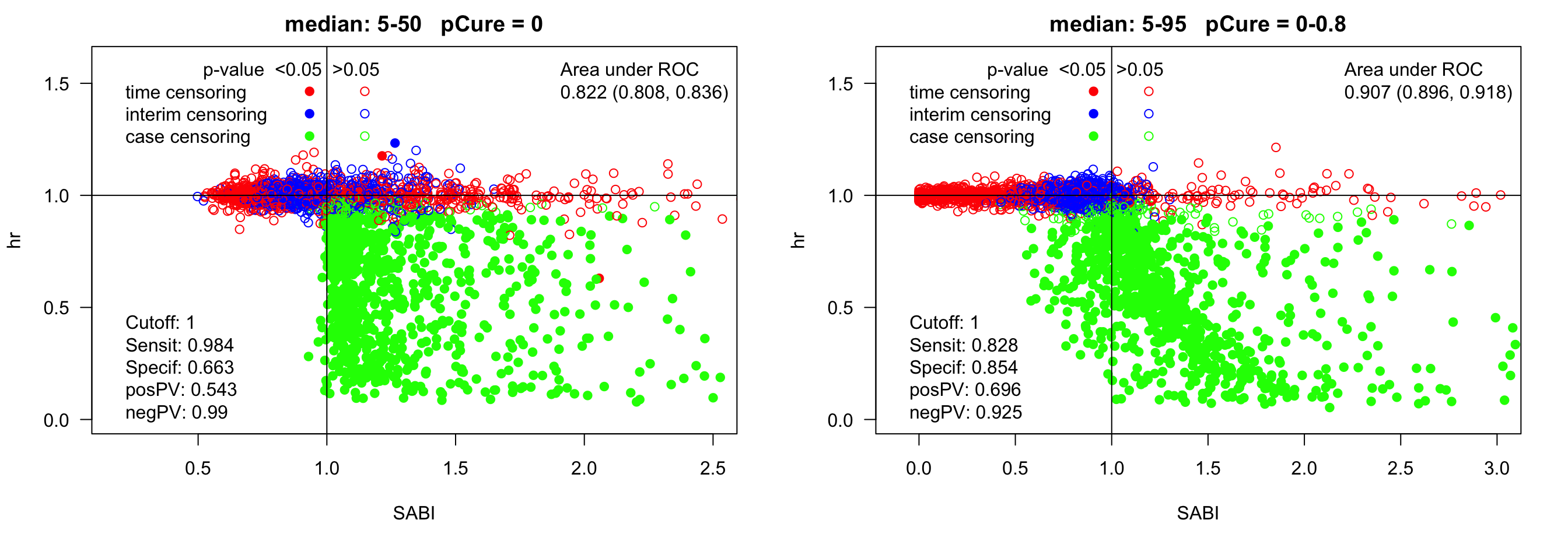}
  \caption{Influence of SABI on survival estimation for right skewed distributions with cure-rate = 0 (left panel) or overall (right panel)}
  \label{fig5}
\end{figure}

After these adjustments, the ABI cutoff for trials based on median event
times of 5 to 50 units and cure-rate of 0 is 0.909, with an area under
the ROC curve of 0.822 (95\% CI 0.808-836), worse than for the QBI.
However, as can ben seen in figure \ref{fig5}, for datasets based on
median event times of 5-95 units and cure rate of 0-80\%, the cutoff is
0.932, and the area under the ROC curve increases to 0.91 (95\% CI:
0.899-0.921). A scaled ABI (SABI) is calculated dividing the ABI by the
cutoff:

\[
  \Large {SABI} = \frac{ ABI } {{cutoff}}
\]

Thus, both scaled indexes, SQBI and SABI, appear effective to detect
bias in the survival analysis of clinical trials and are supplementary.
SQBI appears superior to SABI for right skewed Weibull distributions
with cure rate of 0\%, with a high sensitivity of 0.978 and specificity
of 0.91. For all the remained cases, SABI is better, with and
accceptable sensitivity of 0.866 and specificity of 0.842.

\hypertarget{clinical-available-datasets}{%
\section{Clinical available
datasets}\label{clinical-available-datasets}}

Available clinical datasets were retrieved from Documentation for
package \(survival\) version 2.44-1.1, based on 6 published articles
\cite{survivalDatasets}. Bias indexes SQBI and SABI were applied to the
datasets and to the research arms where available.

\begin{table}[ht]
\centering
\caption{Bias in available clinical datasests} 
\begin{tabular}{lrrlll}
  \hline
trial & n & pCens & SQBI & SABI & reference \\ 
  \hline
Acute Myelogenous Leukemia survival data &  23 & 0.22 & 0.89 & 0.58 & \cite{Miller1997} \\ 
  Bladder Cancer Recurrences - bladder1 & 294 & 0.35 & 0.51 & 0.69 & \cite{Wei1989} \\ 
   - placebo & 128 & 0.31 & 0.66 & 0.61 &  \\ 
   - pyridoxine &  85 & 0.33 & 0.3 & 0.7 &  \\ 
   - thiotepa &  81 & 0.43 & 0.85 & \textbf{1.3} &  \\ 
  Bladder Cancer Recurrences - bladder2 & 178 & 0.37 & 0.6 & 0.78 & \cite{Wei1989} \\ 
   - rx1 & 106 & 0.32 & 0.69 & 0.8 &  \\ 
   - rx2 &  72 & 0.44 & 0.87 & 0.92 &  \\ 
  NCCTG Lung Cancer Data & 228 & 0.28 & 0.79 & 0.73 & \cite{Loprinzi1994} \\ 
  Chemotherapy for Stage B/C colon cancer & 1858 & 0.50 & 0.59 & 0.04 & \cite{Moertel1990} \\ 
   - Observation & 630 & 0.45 & 0.62 & 0.048 &  \\ 
   - Levamisol & 620 & 0.46 & 0.57 & 0.037 &  \\ 
   - Levamisol + 5FU & 608 & 0.60 & 0.62 & 0.041 &  \\ 
  Ovarian Cancer Survival Data &  26 & 0.54 & \textbf{1.1} & 0 & \cite{Edmonson1979} \\ 
   - rx1 &  13 & 0.46 & 1 & 0 &  \\ 
   - rx2 &  13 & 0.62 & \textbf{1.1} & \textbf{1.8} &  \\ 
  Veterans' Administration Lung Cancer Study & 137 & 0.07 & \textbf{1.5} & \textbf{1.4} & \cite{Kalbfleisch1980} \\ 
   - standard &  69 & 0.07 & \textbf{1.8} & 0.8 &  \\ 
   - test &  68 & 0.06 & \textbf{1.8} & \textbf{2} &  \\ 
   \hline
\end{tabular}
\end{table}

A SQBI over 1 was found in the Veterans' Administration Lung Cancer
Study overall, as well as the standard and test arms, suggesting
right-censoring bias. The Veterans' Administration Lung Cancer Study has
a righ skewed distribution of times and a long-term survival rate of 0,
giving strength to this result as a true positive
\cite{Kalbfleisch1980}. The SQBI of 1.1 found in the Ovarian Cancer
Survival Data overall and in the rx2 treatment arm appears confirmed in
the rx2 arm by a SABI of 1.8, but bias may be the result of the small
number of cases \cite{Edmonson1979}. The isolated SABI of 1.3 for the
thiotepa arm of the Bladder Cancer Recurrences dataset may alert us
about the presence of bias, suggesting the need for updating results.

\hypertarget{discussion}{%
\section{Discussion}\label{discussion}}

In this study, we have addressed the problem of the potential bias
associated with right-censoring in clinical trials, a technical problem
beyond the common biases associated with research design and execution,
because it is inherent to the methodology of Kaplan-Meier survival
analysis and may be the source of bias even in spite of a well designed
and executed trial. We developed three different mechanisms of
right-censoring: time censoring, interim censoring and case censoring,
based on common simulation censoring models, the well known interim
analysis performed before a clinical trial has been completed, and the
loss of information because of patients lost to follow-up or withdraw of
consent, respectively. It is shown that one of the three mechanisms of
censoring, case censoring, equivalent to dropouts and lost to follow-up,
is associated with a biased improvement of results compared with
complete follow-up, as demonstrated by a decrease in the HR of censored
group to complete follow-up group below 1 and a p-value less than 0.05.
This impact on HR is inversely correlated to the proportion of censored
cases. Interim censoring, where entry in the study is random and
censoring date is the date of interim analysis or end of the study,
guarantees that censoring and event times are independent
\cite{Krstajic2017}.

Though case censoring may be considered a hypothetical factor in
clinical trials, a systematic review has acknowledged a median
proportion of patients lost to follow-up of 6\% (when reported, as 13\%
of trials do not give this information at all). Different assumptions
about what proportion of cases lost to follow-up end having the event
may change the results in up to a third of trials \cite{Akl2012}. The
most extreme case is the estimated fourfold overestimation of survival
in patients with a resected pancreatic cancer which have been attributed
to a 60\% of patients lost to follow-up \cite{Gudjonsson2002}. A
systematic review of clinical trials on antiretroviral therapy in low
and middle income countries recognized that a large proportion of
patients are lost to follow-up. Increasing physical tracing is
associated with a lower proportion of patients lost to follow-up (7.6\%
vs.~15.1\%; p \textless{} 0.001) and higher estimated mortality (10.5\%
vs.~6.6\%; p = 0.006), in spite of higher retention on treatment (80.0\%
vs.~72.9\%; p = 0.02), suggesting an overestimation of survival
associated with censoring \cite{McMahon2013}. In the same way, the
Kaplan-Meier survival analysis of patients with squamous cell carcinoma
of head and neck reveals that updating the information of patients lost
to follow-up anytime, decreases significantly the 10 years survival
estimation (28.2\% vs 46\%, p \textless{} 0.001) \cite{Priante2005}. The
theoretical framework we present here is consistent with all those
studies.

A first solution to the problem of bias associated with incomplete
follow-up is to maximize event information. Even though patient
participation in a clinical trial may be stopped, at any time, by
patient or clinician decision, withdrawn consent for treatment does not
imply withdrawn consent for obtaining information. It should be
imperative for patients, clinicians and researchers to maintain the loss
of information to a minimum, though data on morbidity may be more
difficult to obtain than mortality data in these patients
\cite{Gabriel2011}. In addition, the censoring proportion should be
reported in the control and experiental arms of clinical trials, as
systematic bias may influence results \cite{Rosen2020}.

The two presented bias indexes, SQBI and SABI, may suggest the presence
of bias and help intensify the efforts to obtain information. In serious
diseases where the event is the norm in a short time interval, such as
advanced pancreatic cancer, SQBI may be appropriate; in more indolent
diseases with long-term survival rate, as the case of adjuvant treatment
in breast cancer patients, SABI may be more appropriate. Aggressive
neoplasms with some long-term survivors, as recent trials of advanced
lung cancer on immunotherapy, represent a challenge for both BI. The
negative predictive values are very high: 0.99 for SQBI and 0.939 for
SABI, suggesting that a BI less than 1 is reassuring. Unfortunately, the
positive predictive values are lower: 0.816 for SQBI and 0.69 por SABI,
and both indexes should be used as a warning of bias, as they are
associated with 20-30\% of false positives. Thus, a SQBI and/or SABI
greater than 1 should warn about potential bias and indicate the need of
updating results and replication. Replication of clinical trials results
using other trials or real-world evidence should be pursued to advance
certainty and regulatory science \cite{Sheffield2020}.

It has been claimed that survival data may be viewed as the competition
of three risks: risc of failure/event, administrative censoring, both
mutually independent, and dropout/loss to follow-up censoring, which may
not be independent of failure risk \cite{Krstajic2017}. A strategy to
attenuate right-censoring bias could be to differentiate lost to
follow-up from other kinds of censoring, and treat them differently in
the Kaplan-Meier analysis, such as assigning them a certain risk of
event in the interval, interpolated from the previous and following
event cases, or modifying its follow-up time by a correction factor.

The bias detected in the Veterans' Administration Lung Cancer Study are
proof of concept that right-censoring bias in clinical trials may be
detected and measured. It would have been very helpful to update the
survival time of censored cases to know if these results hold, as has
been reported in the forementioned squamous cell carcinoma of head and
neck trial \cite{Priante2005}. Censoring bias analysis should be pursued
in clinical trials. A censoring bias analysis in the public platform
\emph{projectdatasphere.org} is under way.

\hfill

\textbf {Acknowledgements}

The authors thank Dr.~A. López, for his constructive criticism and
discussion on this subject, and to the people that create and maintain
open-source software packages we use.

\hfill

\textbf {Conflict of interest}

None

\hfill

\textbf {Data availability}

The data and model code on which these results are based will be
available in the APPENDIX, at the end of this article.

\hfill

\textbf {ORCID}

Enrique Barrajón \url{https://orcid.org/0000-0002-6648-5704} \newline
Laura Barrajón \url{https://orcid.org/0000-0002-2004-3708}

\bibliographystyle{unsrt}
\bibliography{RightCensoring.bib}

\end{document}


\setcounter{page}{13}

\LARGE

\textbf{Appendix}

\vspace{5mm}

\normalsize

\textbf{censoringCode.Rmd}

10 Dec 2020

\hypertarget{initialization}{%
\subsection{Initialization}\label{initialization}}

\begin{Shaded}
\begin{Highlighting}[]
\KeywordTok{rm}\NormalTok{(}\DataTypeTok{list =} \KeywordTok{ls}\NormalTok{())}
\KeywordTok{library}\NormalTok{(}\StringTok{"knitr"}\NormalTok{)}
\KeywordTok{library}\NormalTok{(}\StringTok{"survival"}\NormalTok{)}
\KeywordTok{library}\NormalTok{(}\StringTok{"OptimalCutpoints"}\NormalTok{)}
\KeywordTok{library}\NormalTok{(}\StringTok{"xtable"}\NormalTok{)}
\end{Highlighting}
\end{Shaded}

\hypertarget{survival-functions}{%
\subsection{Survival functions}\label{survival-functions}}

This function takes 2 quantile times and the corresponding quantile
proportions and returns the parameters of a Weibull distribution

@param Q\_A A positive number

@param Q\_B A positive number

@param p\_A A probability \(p_A \in (0, 1)\) corresponding to the
quantile indicated by the time Q\_A

@param p\_B A probability \(p_B \in (0, 1)\) corresponding to the
quantile indicated by the time Q\_B

@return a list The parameters \emph{shape} and \emph{scale} of a Weibull
distribution

\vspace{2mm}

\begin{Shaded}
\begin{Highlighting}[]
\NormalTok{quantiles2weibull <-}\StringTok{ }\ControlFlowTok{function}\NormalTok{(Q_A, Q_B, }\DataTypeTok{p_A =} \FloatTok{0.5}\NormalTok{, }\DataTypeTok{p_B =} \FloatTok{0.99}\NormalTok{) \{}
\NormalTok{   k =}\StringTok{ }\KeywordTok{log}\NormalTok{(}\KeywordTok{log}\NormalTok{(}\DecValTok{1} \OperatorTok{-}\StringTok{ }\NormalTok{p_B) }\OperatorTok{/}\StringTok{ }\KeywordTok{log}\NormalTok{(}\DecValTok{1}\OperatorTok{-}\StringTok{ }\NormalTok{p_A)) }\OperatorTok{/}\StringTok{ }\KeywordTok{log}\NormalTok{(Q_B }\OperatorTok{/}\StringTok{ }\NormalTok{Q_A)}
\NormalTok{   b =}\StringTok{ }\NormalTok{Q_A }\OperatorTok{/}\StringTok{ }\NormalTok{(}\OperatorTok{-}\KeywordTok{log}\NormalTok{(}\DecValTok{1}\OperatorTok{-}\StringTok{ }\NormalTok{p_A))}\OperatorTok{^}\NormalTok{(}\DecValTok{1}\OperatorTok{/}\NormalTok{k)}
   \KeywordTok{list}\NormalTok{(}\DataTypeTok{shape =}\NormalTok{ k, }\DataTypeTok{scale =}\NormalTok{ b)}
\NormalTok{\}}
\end{Highlighting}
\end{Shaded}

\vspace{5mm}

This function takes 2 quantile times and the corresponding quantile
proportions and returns a function that transform a vector of
probabilities into a vector of times following the corresponding Weibull
distribution

@param Q\_A A positive number

@param Q\_B A positive number

@param p\_A A probability \(p_A \in (0, 1)\) corresponding to the
quantile indicated by the time Q\_A

@param p\_B A probability \(p_B \in (0, 1)\) corresponding to the
quantile indicated by the time Q\_B

@param p A vector of probabilities

@return a vector The vector of times

\vspace{2mm}

\begin{Shaded}
\begin{Highlighting}[]
\NormalTok{inverseWeibull <-}\StringTok{ }\ControlFlowTok{function}\NormalTok{(Q_A, Q_B, }\DataTypeTok{p_A =} \FloatTok{0.5}\NormalTok{, }\DataTypeTok{p_B =} \FloatTok{0.99}\NormalTok{) \{}
   \ControlFlowTok{function}\NormalTok{(}\DataTypeTok{p =} \KeywordTok{runif}\NormalTok{(}\DecValTok{1}\NormalTok{)) \{}
\NormalTok{      qWeibull <-}\StringTok{ }\KeywordTok{quantiles2weibull}\NormalTok{(Q_A, Q_B, p_A, p_B)}
      \KeywordTok{with}\NormalTok{(qWeibull, }\KeywordTok{sapply}\NormalTok{(p, }\ControlFlowTok{function}\NormalTok{(p) scale }\OperatorTok{*}\StringTok{ }\NormalTok{(}\OperatorTok{-}\StringTok{ }\KeywordTok{log}\NormalTok{(p))}\OperatorTok{^}\NormalTok{(}\DecValTok{1}\OperatorTok{/}\NormalTok{shape)))}
\NormalTok{   \}}
\NormalTok{\}}
\end{Highlighting}
\end{Shaded}

\vspace{5mm}

This function accepts a number of cases and the parameters of a mixed
model (median, \(Q_{99}\), and cure-rate), estimates the cured cases and
the inverse Weibull event times, and returns a virtual dataset with
complete follow-up to be used by the survival package.

@param nCases The number of cases in the virtual dataset

@param Q\_A A positive number, the median event time

@param Q\_B A positive number, the \(Q_{99}\) event time

@param pCure The cure rate \(pCure \in (0, 1)\) of the mixed model

@param x The label for the dataset

@return the virtual dataset. A data.frame with columns c(time, status,
x)

\vspace{2mm}

\begin{Shaded}
\begin{Highlighting}[]
\NormalTok{completeFA <-}\StringTok{ }\ControlFlowTok{function}\NormalTok{(}\DataTypeTok{nCases =} \DecValTok{10}\NormalTok{, }\DataTypeTok{Q_A =} \DecValTok{25}\NormalTok{, }\DataTypeTok{Q_B =} \DecValTok{100}\NormalTok{, }\DataTypeTok{pCure =} \DecValTok{0}\NormalTok{, }\DataTypeTok{x =} \StringTok{"Uncensored"}\NormalTok{) \{}
\NormalTok{   nEvents <-}\StringTok{ }\KeywordTok{as.integer}\NormalTok{(nCases }\OperatorTok{*}\StringTok{ }\NormalTok{(}\DecValTok{1} \OperatorTok{-}\StringTok{ }\NormalTok{pCure))}
\NormalTok{   time =}\StringTok{ }\KeywordTok{sort}\NormalTok{(}\KeywordTok{inverseWeibull}\NormalTok{(Q_A }\OperatorTok{*}\StringTok{ }\NormalTok{(}\DecValTok{1} \OperatorTok{-}\StringTok{ }\NormalTok{pCure), Q_B)(}\KeywordTok{runif}\NormalTok{(nEvents)))}
\NormalTok{   tmax =}\StringTok{ }\KeywordTok{max}\NormalTok{(time)}
\NormalTok{   time =}\StringTok{ }\KeywordTok{c}\NormalTok{(time, }\KeywordTok{rep}\NormalTok{(tmax, nCases }\OperatorTok{-}\StringTok{ }\NormalTok{nEvents))}
\NormalTok{   status <-}\StringTok{ }\KeywordTok{c}\NormalTok{(}\KeywordTok{rep}\NormalTok{(}\DecValTok{1}\NormalTok{, nEvents), }\KeywordTok{rep}\NormalTok{(}\DecValTok{0}\NormalTok{, nCases }\OperatorTok{-}\StringTok{ }\NormalTok{nEvents))}
   \KeywordTok{data.frame}\NormalTok{(time, status, }\DataTypeTok{x =}\NormalTok{ x, }\DataTypeTok{stringsAsFactors =} \OtherTok{FALSE}\NormalTok{)}
\NormalTok{\}}
\end{Highlighting}
\end{Shaded}

\hypertarget{censoring-functions}{%
\subsection{Censoring functions}\label{censoring-functions}}

This function takes a complete follow-up dataset and the parameters of a
mixed model (events median and \(Q_{99}\), and proportion of censoring),
estimates a vector of random probabilities and the inverse Weibull time,
updates the dataset time and status when the new time is shorter, and
returns a censored dataset

@param uncensored The complete follow-up dataset

@param Q\_A The median time of the distribution

@param Q\_B the \(Q_{99}\) time of the distribution

@param pCensoring A proportion of censored cases

@return a censored dataset. A data.frame with columns c(time, status, x)

\vspace{2mm}

\begin{Shaded}
\begin{Highlighting}[]
\NormalTok{timeCensoring <-}\StringTok{ }\ControlFlowTok{function}\NormalTok{(uncensored, }\DataTypeTok{Q_A =} \DecValTok{25}\NormalTok{, }\DataTypeTok{Q_B =} \DecValTok{100}\NormalTok{, }\DataTypeTok{pCensoring =} \DecValTok{0}\NormalTok{) \{}
\NormalTok{   censored <-}\StringTok{ }\NormalTok{uncensored}
\NormalTok{   n <-}\StringTok{ }\KeywordTok{nrow}\NormalTok{(censored)}
\NormalTok{   pRandom <-}\StringTok{ }\KeywordTok{runif}\NormalTok{(n)}\OperatorTok{^}\NormalTok{((}\DecValTok{1}\OperatorTok{-}\NormalTok{pCensoring)}\OperatorTok{/}\NormalTok{(pCensoring))}
\NormalTok{   timeCens <-}\StringTok{ }\KeywordTok{sapply}\NormalTok{(pRandom, }\KeywordTok{inverseWeibull}\NormalTok{(Q_A, Q_B))}
\NormalTok{   censor <-}\StringTok{ }\NormalTok{timeCens }\OperatorTok{<}\StringTok{ }\NormalTok{censored}\OperatorTok{$}\NormalTok{time}
\NormalTok{   censored}\OperatorTok{$}\NormalTok{time[censor] <-}\StringTok{ }\NormalTok{timeCens[censor]}
\NormalTok{   censored}\OperatorTok{$}\NormalTok{status[censor] <-}\StringTok{ }\DecValTok{0}
\NormalTok{   censored}\OperatorTok{$}\NormalTok{x <-}\StringTok{ }\KeywordTok{paste}\NormalTok{(}\StringTok{"Time censoring"}\NormalTok{, }\KeywordTok{round}\NormalTok{(}\DecValTok{100} \OperatorTok{*}\StringTok{ }\NormalTok{pCensoring, }\DecValTok{1}\NormalTok{), }\StringTok{"
\NormalTok{   censored[}\KeywordTok{order}\NormalTok{(censored}\OperatorTok{$}\NormalTok{time), ]}
\NormalTok{\}}
\end{Highlighting}
\end{Shaded}

\vspace{5mm}

This function takes a complete follow-up dataset and the parameters of a
mixed model (events median and \(Q_{99}\), and proportion of censoring),
estimates a vector of random interim times (from a vector of random
probabilities and the inverse Weibull time) and of random recruit times,
updates the dataset time and status when the difference of recruit and
interim times is shorter than the original time, and returns a censored
dataset

@param uncensored The complete follow-up dataset

@param Q\_A The median time of the distribution

@param Q\_B the \(Q_{99}\) time of the distribution

@param pCensoring A proportion of censored cases

@return a censored dataset. A data.frame with columns c(time, status, x)

\vspace{2mm}

\begin{Shaded}
\begin{Highlighting}[]
\NormalTok{interimCensoring <-}\StringTok{ }\ControlFlowTok{function}\NormalTok{(uncensored, }\DataTypeTok{Q_A =} \DecValTok{25}\NormalTok{, }\DataTypeTok{Q_B =} \DecValTok{100}\NormalTok{, }\DataTypeTok{pCensoring =} \DecValTok{0}\NormalTok{) \{}
\NormalTok{   censored <-}\StringTok{ }\NormalTok{uncensored}
\NormalTok{   n <-}\StringTok{ }\KeywordTok{nrow}\NormalTok{(censored)}
\NormalTok{   interimTime <-}\StringTok{ }\KeywordTok{inverseWeibull}\NormalTok{(Q_A, Q_B)(pCensoring) }\OperatorTok{*}\StringTok{ }\DecValTok{2}
\NormalTok{   recruitTime <-}\StringTok{ }\KeywordTok{runif}\NormalTok{(n, }\DecValTok{0}\NormalTok{, interimTime)}
\NormalTok{   interval <-}\StringTok{ }\NormalTok{interimTime }\OperatorTok{-}\StringTok{ }\NormalTok{recruitTime}
\NormalTok{   censor <-}\StringTok{ }\NormalTok{(interimTime }\OperatorTok{-}\StringTok{ }\NormalTok{recruitTime) }\OperatorTok{<}\StringTok{ }\NormalTok{censored}\OperatorTok{$}\NormalTok{time}
\NormalTok{   censored}\OperatorTok{$}\NormalTok{status[censor] <-}\StringTok{ }\DecValTok{0}
\NormalTok{   censored}\OperatorTok{$}\NormalTok{time[censor] <-}\StringTok{ }\NormalTok{interval[censor]}
\NormalTok{   censored}\OperatorTok{$}\NormalTok{x <-}\StringTok{ }\KeywordTok{paste}\NormalTok{(}\StringTok{"Interim at t ="}\NormalTok{, }\KeywordTok{round}\NormalTok{(interimTime, }\DecValTok{1}\NormalTok{))}
\NormalTok{   censored <-}\StringTok{ }\NormalTok{censored[censored}\OperatorTok{$}\NormalTok{time }\OperatorTok{>}\StringTok{ }\DecValTok{0}\NormalTok{, ]}
\NormalTok{   censored[}\KeywordTok{order}\NormalTok{(censored}\OperatorTok{$}\NormalTok{time), ]}
\NormalTok{\}}
\end{Highlighting}
\end{Shaded}

\vspace{5mm}

This function takes a complete follow-up dataset and a proportion of
censoring, selects a random sample of cases with that probability,
updating the status and times (shortening them a random amount), and
returns a censored dataset

@param uncensored The complete follow-up dataset

@param pCensoring A proportion of censored cases

@return a censored dataset. A data.frame with columns c(time, status, x)

\vspace{2mm}

\begin{Shaded}
\begin{Highlighting}[]
\NormalTok{caseCensoring <-}\StringTok{ }\ControlFlowTok{function}\NormalTok{(group0, }\DataTypeTok{pCensoring =} \DecValTok{0}\NormalTok{) \{}
\NormalTok{   n <-}\StringTok{ }\KeywordTok{nrow}\NormalTok{(group0)}
\NormalTok{   nCens  <-}\StringTok{ }\KeywordTok{as.integer}\NormalTok{(n }\OperatorTok{*}\StringTok{ }\NormalTok{pCensoring)}
\NormalTok{   censor <-}\StringTok{ }\KeywordTok{sample}\NormalTok{(}\DecValTok{1}\OperatorTok{:}\NormalTok{n, nCens) }
\NormalTok{   censored <-}\StringTok{ }\NormalTok{group0}
\NormalTok{   nCensored <-}\StringTok{ }\KeywordTok{as.integer}\NormalTok{(}\KeywordTok{nrow}\NormalTok{(group0) }\OperatorTok{*}\StringTok{ }\NormalTok{pCensoring)}
\NormalTok{   censored}\OperatorTok{$}\NormalTok{time[censor] <-}\StringTok{ }\NormalTok{group0}\OperatorTok{$}\NormalTok{time[censor] }\OperatorTok{*}\StringTok{ }\KeywordTok{runif}\NormalTok{(nCensored, }\FloatTok{0.2}\NormalTok{, }\DecValTok{1}\NormalTok{)}
\NormalTok{   censored}\OperatorTok{$}\NormalTok{status[censor] <-}\StringTok{ }\DecValTok{0}
\NormalTok{   censored}\OperatorTok{$}\NormalTok{x <-}\StringTok{ }\KeywordTok{paste}\NormalTok{(}\StringTok{"Case"}\NormalTok{, }\KeywordTok{round}\NormalTok{(}\DecValTok{100} \OperatorTok{*}\StringTok{ }\NormalTok{(}\DecValTok{1} \OperatorTok{-}\StringTok{ }\NormalTok{pCensoring), }\DecValTok{1}\NormalTok{), }\StringTok{"
\NormalTok{   censored[}\KeywordTok{order}\NormalTok{(censored}\OperatorTok{$}\NormalTok{time), ]}
\NormalTok{\}}
\end{Highlighting}
\end{Shaded}

\hypertarget{bias-indexes-bi}{%
\subsection{Bias indexes (BI)}\label{bias-indexes-bi}}

This function estimates the quantile bias index (QBI) of a dataset

@param dataset The survival dataset with columns c(time, status, x)

@param time The name of the column corresponding to time

@param status The name of the column correspondint to status

@param event The label of status corresponding to censoring

@return a number The QBI

\vspace{2mm}

\begin{Shaded}
\begin{Highlighting}[]
\NormalTok{QBI <-}\StringTok{ }\ControlFlowTok{function}\NormalTok{(dataset, }\DataTypeTok{time =} \StringTok{"time"}\NormalTok{, }\DataTypeTok{status =} \StringTok{"status"}\NormalTok{, }\DataTypeTok{event =} \DecValTok{1}\NormalTok{, ...) \{}
\NormalTok{   events <-}\StringTok{ }\NormalTok{dataset[dataset[, status] }\OperatorTok{==}\StringTok{ }\NormalTok{event, time]}
\NormalTok{   censor <-}\StringTok{ }\NormalTok{dataset[dataset[, status] }\OperatorTok{!=}\StringTok{ }\NormalTok{event, time]}
\NormalTok{   aCensor <-}\StringTok{ }\NormalTok{censor[censor }\OperatorTok{<}\StringTok{ }\KeywordTok{max}\NormalTok{(events)]}
   \ControlFlowTok{if}\NormalTok{(}\KeywordTok{length}\NormalTok{(aCensor) }\OperatorTok{==}\StringTok{ }\DecValTok{0}\NormalTok{) }\OtherTok{NA} \ControlFlowTok{else} \KeywordTok{quantile}\NormalTok{(events, }\FloatTok{0.95}\NormalTok{) }\OperatorTok{/}\StringTok{ }\KeywordTok{quantile}\NormalTok{(aCensor, }\FloatTok{0.95}\NormalTok{)}
\NormalTok{\}}
\end{Highlighting}
\end{Shaded}

\vspace{5mm}

This function estimates the scaled quantile bias index (SQBI) of a
dataset

@param dataset The survival dataset with columns c(time, status, x)

@param time The name of the column corresponding to time

@param status The name of the column correspondint to status

@param event The label of status corresponding to censoring

@return a number The SQBI

\vspace{2mm}

\begin{Shaded}
\begin{Highlighting}[]
\NormalTok{SQBI <-}\StringTok{ }\ControlFlowTok{function}\NormalTok{(dataset, }\DataTypeTok{time =} \StringTok{"time"}\NormalTok{, }\DataTypeTok{status =} \StringTok{"status"}\NormalTok{, }\DataTypeTok{event =} \DecValTok{1}\NormalTok{, ...) \{}
   \KeywordTok{QBI}\NormalTok{(dataset, time, status, event) }\OperatorTok{/}\StringTok{ }\FloatTok{1.2}
\NormalTok{\}}
\end{Highlighting}
\end{Shaded}

\vspace{5mm}

This function estimates the under the mean bias index (UMBI) of a
dataset

@param dataset The survival dataset with columns c(time, status, x)

@param time The name of the column corresponding to time

@param status The name of the column correspondint to status

@param event The label of status corresponding to censoring

@return a number The UMBI

\vspace{2mm}

\begin{Shaded}
\begin{Highlighting}[]
\NormalTok{UMBI <-}\StringTok{ }\ControlFlowTok{function}\NormalTok{(dataset, }\DataTypeTok{time =} \StringTok{"time"}\NormalTok{, }\DataTypeTok{status =} \StringTok{"status"}\NormalTok{, }\DataTypeTok{event =} \DecValTok{1}\NormalTok{, ...) \{}
\NormalTok{   events <-}\StringTok{ }\NormalTok{dataset[dataset[, status] }\OperatorTok{==}\StringTok{ }\NormalTok{event, time]}
\NormalTok{   censor <-}\StringTok{ }\NormalTok{dataset[dataset[, status] }\OperatorTok{!=}\StringTok{ }\NormalTok{event, time]}
\NormalTok{   lastEvent <-}\StringTok{ }\KeywordTok{max}\NormalTok{(events)}
\NormalTok{   meanEvent <-}\StringTok{ }\KeywordTok{mean}\NormalTok{(events)}
\NormalTok{   aCensor <-}\StringTok{ }\NormalTok{censor[censor }\OperatorTok{<}\StringTok{ }\NormalTok{lastEvent]}
   \KeywordTok{sum}\NormalTok{(aCensor }\OperatorTok{<}\StringTok{ }\NormalTok{meanEvent) }\OperatorTok{/}\StringTok{ }\KeywordTok{length}\NormalTok{(aCensor)}
\NormalTok{\}}
\end{Highlighting}
\end{Shaded}

\vspace{5mm}

This function estimates the adjusted bias index (ABI) of a dataset

@param dataset The survival dataset with columns c(time, status, x)

@param time The name of the column corresponding to time

@param status The name of the column correspondint to status

@param event The label of status corresponding to censoring

@return a number The ABI

\vspace{2mm}

\begin{Shaded}
\begin{Highlighting}[]
\NormalTok{ABI <-}\StringTok{ }\ControlFlowTok{function}\NormalTok{(dataset, }\DataTypeTok{time =} \StringTok{"time"}\NormalTok{, }\DataTypeTok{status =} \StringTok{"status"}\NormalTok{, }\DataTypeTok{event =} \DecValTok{1}\NormalTok{, ...) \{}
\NormalTok{   events <-}\StringTok{ }\NormalTok{dataset[dataset[, status] }\OperatorTok{==}\StringTok{ }\NormalTok{event, time]}
\NormalTok{   censor <-}\StringTok{ }\NormalTok{dataset[dataset[, status] }\OperatorTok{!=}\StringTok{ }\NormalTok{event, time]}
\NormalTok{   lastEvent <-}\StringTok{ }\KeywordTok{max}\NormalTok{(events)}
\NormalTok{   meanEvent <-}\StringTok{ }\KeywordTok{mean}\NormalTok{(events)}
\NormalTok{   aCensor<-}\StringTok{ }\NormalTok{censor[censor }\OperatorTok{<}\StringTok{ }\NormalTok{lastEvent]}
\NormalTok{   formula1 <-}\StringTok{ }\KeywordTok{Surv}\NormalTok{(time, status }\OperatorTok{==}\StringTok{ }\NormalTok{event) }\OperatorTok{~}\StringTok{ }\DecValTok{1}
\NormalTok{   survfit1 <-}\StringTok{ }\KeywordTok{survfit}\NormalTok{(formula1, }\DataTypeTok{data =}\NormalTok{ dataset)}
\NormalTok{   pLongTerm <-}\StringTok{ }\KeywordTok{summary}\NormalTok{(survfit1, }\DataTypeTok{times =}\NormalTok{ lastEvent)}\OperatorTok{$}\NormalTok{surv}
\NormalTok{   medianCorrection <-}\StringTok{ }\KeywordTok{mean}\NormalTok{(events) }\OperatorTok{/}\StringTok{ }\KeywordTok{median}\NormalTok{(events)}
\NormalTok{   longTermCorrection <-}\StringTok{ }\KeywordTok{exp}\NormalTok{(pLongTerm)}
   \KeywordTok{sum}\NormalTok{(aCensor }\OperatorTok{<}\StringTok{ }\NormalTok{meanEvent) }\OperatorTok{/}\StringTok{ }\KeywordTok{length}\NormalTok{(aCensor) }\OperatorTok{*}\StringTok{ }\NormalTok{medianCorrection }\OperatorTok{*}\StringTok{ }\NormalTok{longTermCorrection}
\NormalTok{\}}
\end{Highlighting}
\end{Shaded}

\vspace{5mm}

This function estimates the scaled adjusted bias index (ABI) of a
dataset

@param dataset The survival dataset with columns c(time, status, x)

@param time The name of the column corresponding to time

@param status The name of the column correspondint to status

@param event The label of status corresponding to censoring

@return

\vspace{2mm}

\begin{Shaded}
\begin{Highlighting}[]
\NormalTok{SABI <-}\StringTok{ }\ControlFlowTok{function}\NormalTok{(dataset, }\DataTypeTok{time =} \StringTok{"time"}\NormalTok{, }\DataTypeTok{status =} \StringTok{"status"}\NormalTok{, }\DataTypeTok{event =} \DecValTok{1}\NormalTok{, ...) \{}
   \KeywordTok{ABI}\NormalTok{(dataset, time, status, event) }\OperatorTok{/}\StringTok{ }\FloatTok{0.932}
\NormalTok{\}}
\end{Highlighting}
\end{Shaded}

\hypertarget{experiment-functions}{%
\subsection{Experiment functions}\label{experiment-functions}}

This function extracts the hazard ratio and p-value between the two
groups labeled in the x column of a survival dataset

@param dataset The survival dataset with columns c(time, status, x)

@param time The name of the column corresponding to time

@param status The name of the column correspondint to status

@param event The label of status corresponding to censoring

@param x Labels for the two groups in the dataset

@return a list with the hazard ratio and p-value betwen the two groups

\vspace{2mm}

\begin{Shaded}
\begin{Highlighting}[]
\NormalTok{getHR <-}\StringTok{ }\ControlFlowTok{function}\NormalTok{(dataset, }\DataTypeTok{time =} \StringTok{"time"}\NormalTok{, }\DataTypeTok{status =} \StringTok{"status"}\NormalTok{, }\DataTypeTok{event =} \DecValTok{1}\NormalTok{, }\DataTypeTok{x =} \StringTok{"x"}\NormalTok{) \{}
\NormalTok{   formula1 =}\StringTok{ }\KeywordTok{Surv}\NormalTok{(time, status }\OperatorTok{==}\StringTok{ }\NormalTok{event) }\OperatorTok{~}\StringTok{ }\NormalTok{x}
\NormalTok{   cx <-}\StringTok{ }\KeywordTok{summary}\NormalTok{(}\KeywordTok{suppressWarnings}\NormalTok{(}\KeywordTok{coxph}\NormalTok{(formula1, }\DataTypeTok{data =}\NormalTok{ dataset)))}
   \KeywordTok{list}\NormalTok{(}\DataTypeTok{hr =}\NormalTok{ cx}\OperatorTok{$}\NormalTok{conf.int[[}\DecValTok{1}\NormalTok{]], }\DataTypeTok{p =}\NormalTok{ cx}\OperatorTok{$}\NormalTok{coefficients[[}\DecValTok{5}\NormalTok{]])}
\NormalTok{\}}
\end{Highlighting}
\end{Shaded}

\vspace{5mm}

This function takes a complete follow-up and a censored datasets, the
names of columns corresponding to time and status, and the label of
status corresponding to censoring, and a BI function, returning a vector
with the parameters of the trial

@param group0

@param group1

@param time The name of the column corresponding to time

@param status The name of the column correspondint to status

@param event The label of status corresponding to censoring

@return a vector The parameters of the trial: number of cases,
proportion of long-term survivors, hazard ratio, p-value and BI result

\vspace{2mm}

\begin{Shaded}
\begin{Highlighting}[]
\NormalTok{trialResults <-}
\StringTok{   }\ControlFlowTok{function}\NormalTok{(group0, group1, }\DataTypeTok{time =} \StringTok{"time"}\NormalTok{, }\DataTypeTok{status =} \StringTok{"status"}\NormalTok{, }\DataTypeTok{event =} \DecValTok{1}\NormalTok{) \{}
\NormalTok{      events <-}\StringTok{ }\NormalTok{group1[group1[, status] }\OperatorTok{==}\StringTok{ }\NormalTok{event, time]}
\NormalTok{      lastEvent <-}\StringTok{ }\KeywordTok{max}\NormalTok{(events)}
\NormalTok{      aCensor <-}\StringTok{ }\NormalTok{group1[group1[, time] }\OperatorTok{<}\StringTok{ }\NormalTok{lastEvent, ]}
\NormalTok{      pCensor <-}\StringTok{ }\KeywordTok{sum}\NormalTok{(aCensor[, status] }\OperatorTok{!=}\StringTok{ }\NormalTok{event) }\OperatorTok{/}\StringTok{ }\KeywordTok{nrow}\NormalTok{(aCensor)}
\NormalTok{      pLongTerm <-}\StringTok{ }\DecValTok{1} \OperatorTok{-}\StringTok{ }\KeywordTok{nrow}\NormalTok{(aCensor) }\OperatorTok{/}\StringTok{ }\KeywordTok{nrow}\NormalTok{(group1)}
\NormalTok{      trialDataset <-}\StringTok{ }\KeywordTok{rbind}\NormalTok{(group1, group0)}
\NormalTok{      trialDataset}\OperatorTok{$}\NormalTok{x <-}\StringTok{ }\KeywordTok{factor}\NormalTok{(trialDataset}\OperatorTok{$}\NormalTok{x, }\DataTypeTok{levels =} \KeywordTok{c}\NormalTok{(group0}\OperatorTok{$}\NormalTok{x[[}\DecValTok{1}\NormalTok{]], group1}\OperatorTok{$}\NormalTok{x[[}\DecValTok{1}\NormalTok{]]))}
\NormalTok{      cox <-}\StringTok{ }\KeywordTok{getHR}\NormalTok{(trialDataset)}
      \KeywordTok{c}\NormalTok{(}\KeywordTok{nrow}\NormalTok{(group1), pCensor, pLongTerm, cox}\OperatorTok{$}\NormalTok{hr, cox}\OperatorTok{$}\NormalTok{p,}
        \KeywordTok{SQBI}\NormalTok{(group1), }\KeywordTok{UMBI}\NormalTok{(group1), }\KeywordTok{SABI}\NormalTok{(group1))}
\NormalTok{   \}}
\end{Highlighting}
\end{Shaded}

\vspace{5mm}

This function simulates a number of virtual trials with the indicated
conditions and returns a data.frame with the result of the trial. Every
row is a virtual trial and columns are type of censoring, number of
cases, proportion of censoring, proportion of long-term survivors,
hazard ratio, p-value and BI result

@param nTrials The number of virtual trials to be simulated

@param nCases The number of cases in every virtual trial

@param mediana The median time of events in the censoring dataset

@param pCure The cure-rate in the complete follow-up dataset

@param pCensoring The proportion of censored cases

@param bias The bias function

@return a data.frame

\vspace{2mm}

\begin{Shaded}
\begin{Highlighting}[]
\NormalTok{simulTrials <-}\StringTok{ }\ControlFlowTok{function}\NormalTok{(}\DataTypeTok{nTrials =} \DecValTok{300}\NormalTok{, }\DataTypeTok{nCases =} \DecValTok{1000}\NormalTok{, }\DataTypeTok{mediana =} \DecValTok{20}\NormalTok{, }\DataTypeTok{pCure =} \DecValTok{0}\NormalTok{,}
                        \DataTypeTok{pCensoring =} \StringTok{"runif(1, 0.05, 0.95)"}\NormalTok{, ...) \{}
\NormalTok{   df <-}\StringTok{ }\KeywordTok{data.frame}\NormalTok{(}\KeywordTok{matrix}\NormalTok{(}\DataTypeTok{nrow =} \DecValTok{3} \OperatorTok{*}\StringTok{ }\NormalTok{nTrials, }\DataTypeTok{ncol =} \DecValTok{9}\NormalTok{), }\DataTypeTok{stringsAsFactors =} \OtherTok{FALSE}\NormalTok{)}
   \KeywordTok{colnames}\NormalTok{(df) <-}
\StringTok{      }\KeywordTok{c}\NormalTok{(}\StringTok{"type"}\NormalTok{, }\StringTok{"nCases"}\NormalTok{, }\StringTok{"pCensored"}\NormalTok{, }\StringTok{"pCured"}\NormalTok{, }\StringTok{"hr"}\NormalTok{, }\StringTok{"pValue"}\NormalTok{, }\StringTok{"SQBI"}\NormalTok{, }\StringTok{"UMBI"}\NormalTok{, }\StringTok{"SABI"}\NormalTok{)}
   \ControlFlowTok{for}\NormalTok{(i }\ControlFlowTok{in} \DecValTok{1}\OperatorTok{:}\NormalTok{nTrials) \{}
\NormalTok{      nCases1  <-}\StringTok{ }\KeywordTok{eval}\NormalTok{(}\KeywordTok{parse}\NormalTok{(}\DataTypeTok{text =}\NormalTok{ nCases))}
\NormalTok{      median1  <-}\StringTok{ }\KeywordTok{eval}\NormalTok{(}\KeywordTok{parse}\NormalTok{(}\DataTypeTok{text =}\NormalTok{ mediana))}
\NormalTok{      pCure1   <-}\StringTok{ }\KeywordTok{eval}\NormalTok{(}\KeywordTok{parse}\NormalTok{(}\DataTypeTok{text =}\NormalTok{ pCure))}
\NormalTok{      pCensor1 <-}\StringTok{ }\KeywordTok{eval}\NormalTok{(}\KeywordTok{parse}\NormalTok{(}\DataTypeTok{text =}\NormalTok{ pCensoring))}
\NormalTok{      group0 <-}\StringTok{ }\KeywordTok{completeFA}\NormalTok{(}\DataTypeTok{nCases =}\NormalTok{ nCases1, median1, }\DecValTok{100}\NormalTok{, }\DataTypeTok{pCure =}\NormalTok{ pCure1)}
\NormalTok{      timeCensored    <-}\StringTok{ }\KeywordTok{timeCensoring}\NormalTok{   (group0, }\DataTypeTok{Q_A =}\NormalTok{ median1, }\DecValTok{100}\NormalTok{, }\DataTypeTok{pCensoring =}\NormalTok{ pCensor1)}
\NormalTok{      interimCensored <-}\StringTok{ }\KeywordTok{interimCensoring}\NormalTok{(group0, }\DataTypeTok{Q_A =}\NormalTok{ median1, }\DecValTok{100}\NormalTok{, }\DataTypeTok{pCensoring =}\NormalTok{ pCensor1)}
\NormalTok{      caseCensored    <-}\StringTok{ }\KeywordTok{caseCensoring}\NormalTok{   (group0, }\DataTypeTok{pCensoring =}\NormalTok{ pCensor1)}
\NormalTok{      df[}\DecValTok{3} \OperatorTok{*}\StringTok{ }\NormalTok{i }\OperatorTok{-}\StringTok{ }\DecValTok{2}\OperatorTok{:}\DecValTok{0}\NormalTok{, }\StringTok{"type"}\NormalTok{] <-}\StringTok{ }\KeywordTok{c}\NormalTok{(}\StringTok{"time"}\NormalTok{, }\StringTok{"interim"}\NormalTok{, }\StringTok{"case"}\NormalTok{)}
\NormalTok{      df[}\DecValTok{3} \OperatorTok{*}\StringTok{ }\NormalTok{i }\OperatorTok{-}\StringTok{ }\DecValTok{2}\OperatorTok{:}\DecValTok{0}\NormalTok{, }\DecValTok{-1}\NormalTok{] <-}\StringTok{ }\KeywordTok{rbind}\NormalTok{(}\KeywordTok{trialResults}\NormalTok{(group0, timeCensored),}
                                   \KeywordTok{trialResults}\NormalTok{(group0, interimCensored),}
                                   \KeywordTok{trialResults}\NormalTok{(group0, caseCensored))}
\NormalTok{   \}}
\NormalTok{   df}
\NormalTok{\}}
\end{Highlighting}
\end{Shaded}

\vspace{5mm}

This takes a cancer clinical dataset and returns the row of a data.frame
with columns: name of the trial, number of cases, proportion of
censoring, SQBI and SABI indexes and reference

@param db The clinical cancer dataset

@param trial The name of the trial

@param reference The bibliographic reference

@param time The name of the column corresponding to time

@param status The name of the column correspondint to status

@param event The label of status corresponding to censoring

@return a 1 row data.frame

\vspace{2mm}

\begin{Shaded}
\begin{Highlighting}[]
\NormalTok{clinicalBias <-}\StringTok{ }\ControlFlowTok{function}\NormalTok{(db, trial, }\DataTypeTok{reference =} \StringTok{""}\NormalTok{, }\DataTypeTok{time =} \StringTok{"time"}\NormalTok{, }\DataTypeTok{status =} \StringTok{"status"}\NormalTok{, }\DataTypeTok{event =} \DecValTok{1}\NormalTok{) \{}
\NormalTok{   events    <-}\StringTok{ }\NormalTok{db[, status] }\OperatorTok{==}\StringTok{ }\NormalTok{event}
\NormalTok{   censor    <-}\StringTok{ }\NormalTok{db[, status] }\OperatorTok{!=}\StringTok{ }\NormalTok{event}
\NormalTok{   lastEvent <-}\StringTok{ }\KeywordTok{max}\NormalTok{(db[events, time], }\DataTypeTok{na.rm =} \OtherTok{FALSE}\NormalTok{)}
\NormalTok{   bi1 <-}\StringTok{ }\KeywordTok{SQBI}\NormalTok{(db, }\DataTypeTok{time =}\NormalTok{ time, }\DataTypeTok{status =}\NormalTok{ status, }\DataTypeTok{event =}\NormalTok{ event)}
\NormalTok{   bi3 <-}\StringTok{ }\KeywordTok{SABI}\NormalTok{(db, }\DataTypeTok{time =}\NormalTok{ time, }\DataTypeTok{status =}\NormalTok{ status, }\DataTypeTok{event =}\NormalTok{ event)}
   \KeywordTok{data.frame}\NormalTok{(}\DataTypeTok{trial =}\NormalTok{ trial,}
              \DataTypeTok{n =} \KeywordTok{nrow}\NormalTok{(db),}
              \DataTypeTok{pCens =} \KeywordTok{sum}\NormalTok{(censor) }\OperatorTok{/}\StringTok{ }\KeywordTok{nrow}\NormalTok{(db),}
              \DataTypeTok{SQBI =}\NormalTok{ bi1,}
              \DataTypeTok{SABI =}\NormalTok{ bi3,}
              \DataTypeTok{reference =} \ControlFlowTok{if}\NormalTok{(reference }\OperatorTok{==}\StringTok{ ""}\NormalTok{) }\StringTok{""} \ControlFlowTok{else} \KeywordTok{paste0}\NormalTok{(}\StringTok{"}\CharTok{\textbackslash{}\textbackslash{}}\StringTok{cite\{"}\NormalTok{, reference, }\StringTok{"\}"}\NormalTok{),}
              \DataTypeTok{row.names =} \StringTok{""}\NormalTok{, }\DataTypeTok{stringsAsFactors =} \OtherTok{FALSE}\NormalTok{)}
\NormalTok{\}}
\end{Highlighting}
\end{Shaded}

\hypertarget{plot-functions}{%
\subsection{Plot functions}\label{plot-functions}}

This function plots a Kaplan-Meier curve from a survival dataset. If the
logical newplot is TRUE, the curve is plotted from scratch, otherwise,
the curve is added to the plot

@param dataset The survival dataset with columns c(time, status, x)

@param time The name of the column corresponding to time

@param status The name of the column correspondint to status

@param event The label of status corresponding to censoring

@param x Labels for the groups in the dataset

@param newplot A logical if the curve is to be drawn from scratch or
added

@param col The color for the curve

@return a plot A Kaplan-Meier curve with the median

\vspace{2mm}

\begin{Shaded}
\begin{Highlighting}[]
\NormalTok{KMplot <-}\StringTok{ }\ControlFlowTok{function}\NormalTok{(dataset, }\DataTypeTok{time =} \StringTok{"time"}\NormalTok{, }\DataTypeTok{status =} \StringTok{"status"}\NormalTok{, }\DataTypeTok{event =} \DecValTok{1}\NormalTok{, }\DataTypeTok{x =} \StringTok{"x"}\NormalTok{,}
                   \DataTypeTok{newplot =} \OtherTok{FALSE}\NormalTok{, }\DataTypeTok{col =} \StringTok{"black"}\NormalTok{, ...) \{}
\NormalTok{   op <-}\StringTok{ }\KeywordTok{par}\NormalTok{(}\DataTypeTok{mar =} \KeywordTok{c}\NormalTok{(}\DecValTok{3}\NormalTok{, }\DecValTok{4}\NormalTok{, }\DecValTok{4}\NormalTok{, }\FloatTok{0.8}\NormalTok{), }\DataTypeTok{mgp =} \KeywordTok{c}\NormalTok{(}\DecValTok{2}\NormalTok{, }\FloatTok{0.6}\NormalTok{, }\DecValTok{0}\NormalTok{))}
      \ControlFlowTok{if}\NormalTok{(newplot) \{}
\NormalTok{         xlim =}\StringTok{ }\KeywordTok{c}\NormalTok{(}\DecValTok{0}\NormalTok{, }\KeywordTok{max}\NormalTok{(dataset}\OperatorTok{$}\NormalTok{time, }\DataTypeTok{na.rm =} \OtherTok{FALSE}\NormalTok{) }\OperatorTok{*}\StringTok{ }\FloatTok{1.1}\NormalTok{)}
         \KeywordTok{plot}\NormalTok{(}\OtherTok{NA}\NormalTok{, }\DataTypeTok{type=}\StringTok{"n"}\NormalTok{, }\DataTypeTok{xlab =} \StringTok{"Time"}\NormalTok{, }\DataTypeTok{ylab =} \StringTok{"Proportion"}\NormalTok{, }\DataTypeTok{las =} \DecValTok{1}\NormalTok{,}
              \DataTypeTok{xaxs =} \StringTok{"i"}\NormalTok{, }\DataTypeTok{yaxs =} \StringTok{"i"}\NormalTok{, }\DataTypeTok{xlim =}\NormalTok{ xlim, }\DataTypeTok{ylim =} \KeywordTok{c}\NormalTok{(}\FloatTok{1e-2}\NormalTok{, }\DecValTok{1}\NormalTok{), ...)}
         \KeywordTok{abline}\NormalTok{(}\DataTypeTok{h =} \DecValTok{1}\OperatorTok{:}\DecValTok{3} \OperatorTok{*}\StringTok{ }\FloatTok{0.25}\NormalTok{, }\DataTypeTok{lty =} \StringTok{"dotted"}\NormalTok{)}
\NormalTok{      \}}
\NormalTok{      formula1 <-}\StringTok{ }\KeywordTok{Surv}\NormalTok{(time, status }\OperatorTok{==}\StringTok{ }\NormalTok{event) }\OperatorTok{~}\StringTok{ }\NormalTok{x}
\NormalTok{      survfit1 <-}\StringTok{ }\KeywordTok{survfit}\NormalTok{(formula1, }\DataTypeTok{data =}\NormalTok{ dataset)}
\NormalTok{      median1  <-}\StringTok{ }\KeywordTok{summary}\NormalTok{(survfit1)}\OperatorTok{$}\NormalTok{table[}\StringTok{"median"}\NormalTok{]}
      \KeywordTok{lines}\NormalTok{(survfit1, }\DataTypeTok{mark.time =} \OtherTok{TRUE}\NormalTok{, }\DataTypeTok{conf.int =} \OtherTok{FALSE}\NormalTok{, }\DataTypeTok{col =}\NormalTok{ col, ...)}
      \KeywordTok{lines}\NormalTok{(}\KeywordTok{c}\NormalTok{(median1, median1), }\KeywordTok{c}\NormalTok{(}\OperatorTok{-}\DecValTok{1}\NormalTok{, }\FloatTok{0.5}\NormalTok{), }\DataTypeTok{lty =} \StringTok{"dotted"}\NormalTok{, }\DataTypeTok{lwd =} \DecValTok{2}\NormalTok{, }\DataTypeTok{col =}\NormalTok{ col)}
   \KeywordTok{par}\NormalTok{(op)}
\NormalTok{\}}
\end{Highlighting}
\end{Shaded}

\vspace{5mm}

This function takes the complete follow-up and censored datasets and
calls KMplot to plot the Kaplan-Meier curves. The censored dataset is
plotted in black and the censored dataset in color (red for \emph{time
censoring}, blue for \emph{interim censoring}, and green for \emph{case
censoring}). In addition, the censored dataset is split into two
datasets: events dataset (plotted in orange), and censored in the events
interval (plotted in purple). The hazard ratio and p-value comparing the
two datasets are written in the title after a label

@param group0 The complete follow-up dataset

@param group1 The censored follow-up dataset

@param time The name of the column corresponding to time

@param status The name of the column correspondint to status

@param event The label of status corresponding to censoring

@param x Labels for the groups in the dataset

@param newplot A logical if the curve is to be drawn from scratch or
added

@param main Is a label to compose the plot title

@param col The color for the censored curve

@return a plot The Kaplan-Meier curves of complete follow-up and
censored datasets

\vspace{2mm}

\begin{Shaded}
\begin{Highlighting}[]
\NormalTok{trialPlot <-}\StringTok{ }\ControlFlowTok{function}\NormalTok{(group0, group1, }\DataTypeTok{time =} \StringTok{"time"}\NormalTok{, }\DataTypeTok{status =} \StringTok{"status"}\NormalTok{, }\DataTypeTok{event =} \DecValTok{1}\NormalTok{, }\DataTypeTok{x =} \StringTok{"x"}\NormalTok{,}
                      \DataTypeTok{main =} \StringTok{""}\NormalTok{, }\DataTypeTok{col =} \StringTok{"black"}\NormalTok{, ...) \{}
   \KeywordTok{KMplot}\NormalTok{(group0, }\DataTypeTok{newplot =} \OtherTok{TRUE}\NormalTok{, }\DataTypeTok{lwd =} \DecValTok{2}\NormalTok{)}
   \KeywordTok{KMplot}\NormalTok{(group1, }\DataTypeTok{col =}\NormalTok{ col, ...)}
\NormalTok{   lastEvent <-}\StringTok{ }\KeywordTok{max}\NormalTok{(group0[group0}\OperatorTok{$}\NormalTok{status }\OperatorTok{==}\StringTok{ }\NormalTok{event, time])}
\NormalTok{   events <-}\StringTok{ }\NormalTok{group1[group1[, status] }\OperatorTok{==}\StringTok{ }\NormalTok{event, ]}
\NormalTok{   censor <-}\StringTok{ }\NormalTok{group1[group1[, status] }\OperatorTok{!=}\StringTok{ }\NormalTok{event, ]}
\NormalTok{   censor}\OperatorTok{$}\NormalTok{status <-}\StringTok{ }\DecValTok{1}
   \KeywordTok{KMplot}\NormalTok{(events, }\DataTypeTok{col =} \StringTok{"darkorange"}\NormalTok{, }\DataTypeTok{lwd =} \DecValTok{2}\NormalTok{)}
\NormalTok{   aCensor<-}\StringTok{ }\NormalTok{censor[censor[, time] }\OperatorTok{<}\StringTok{ }\NormalTok{lastEvent, ]}
   \KeywordTok{KMplot}\NormalTok{(aCensor, }\DataTypeTok{col =} \StringTok{"purple"}\NormalTok{, }\DataTypeTok{lwd =} \DecValTok{2}\NormalTok{)}
\NormalTok{   censUncens <-}\StringTok{ }\KeywordTok{rbind}\NormalTok{(group1, group0)}
\NormalTok{   censUncens}\OperatorTok{$}\NormalTok{x <-}\StringTok{ }\KeywordTok{factor}\NormalTok{(censUncens}\OperatorTok{$}\NormalTok{x, }\DataTypeTok{levels =} \KeywordTok{c}\NormalTok{(group0}\OperatorTok{$}\NormalTok{x[[}\DecValTok{1}\NormalTok{]], group1}\OperatorTok{$}\NormalTok{x[[}\DecValTok{1}\NormalTok{]]))}
\NormalTok{   cox <-}\StringTok{ }\KeywordTok{lapply}\NormalTok{(}\KeywordTok{getHR}\NormalTok{(censUncens), signif, }\DataTypeTok{digits =} \DecValTok{3}\NormalTok{)}
   \KeywordTok{title}\NormalTok{(}\DataTypeTok{main =} \KeywordTok{paste0}\NormalTok{(main, }\StringTok{",    hr: "}\NormalTok{, cox}\OperatorTok{$}\NormalTok{hr, }\StringTok{"    p: "}\NormalTok{, cox}\OperatorTok{$}\NormalTok{p), ...)}
\NormalTok{   subgroups <-}\StringTok{ }\KeywordTok{c}\NormalTok{(}\StringTok{"uncensored dataset"}\NormalTok{, }\StringTok{"censored dataset"}\NormalTok{, }\StringTok{"events"}\NormalTok{, }\StringTok{"censored cases"}\NormalTok{)}
   \KeywordTok{legend}\NormalTok{(}\StringTok{"topright"}\NormalTok{, }\DataTypeTok{bty =} \StringTok{"n"}\NormalTok{, }\DataTypeTok{lwd =} \DecValTok{2}\NormalTok{, }\DataTypeTok{xpd=}\OtherTok{TRUE}\NormalTok{,  }\DataTypeTok{legend =}\NormalTok{ subgroups,}
          \DataTypeTok{col =} \KeywordTok{c}\NormalTok{(}\StringTok{"black"}\NormalTok{, col, }\StringTok{"darkorange"}\NormalTok{, }\StringTok{"purple"}\NormalTok{) )}
\NormalTok{\}}
\end{Highlighting}
\end{Shaded}

\vspace{5mm}

This function plots a set of 3 virtual trials from a complete follow-up
dataset and the three types of censoring with the indicated parameters

@param nCases The number of cases in every dataset

@param pCensoring The proportion of censoring in the censoring dataset

@param Q99 The quantile 99 of the events distribution

@param pCure The proportion of cured/long-term survivors

@param mediana The median of the events distribution

@return 3 plots Calls trialPlot 3 times

\vspace{2mm}

\begin{Shaded}
\begin{Highlighting}[]
\NormalTok{virtualTrials <-}\StringTok{ }\ControlFlowTok{function}\NormalTok{(}\DataTypeTok{nCases =} \DecValTok{1000}\NormalTok{, }\DataTypeTok{pCensoring =} \FloatTok{0.5}\NormalTok{, }\DataTypeTok{Q99 =} \DecValTok{100}\NormalTok{, }\DataTypeTok{pCure =} \FloatTok{0.5}\NormalTok{, }\DataTypeTok{mediana =} \DecValTok{25}\NormalTok{) \{}
\NormalTok{      group0 <-}\StringTok{ }\KeywordTok{completeFA}\NormalTok{(nCases, mediana, Q99, }\DataTypeTok{pCure =}\NormalTok{ pCure)}
\NormalTok{      group1 <-}\StringTok{ }\KeywordTok{timeCensoring}\NormalTok{(group0, mediana, Q99, }\DataTypeTok{pCensoring =}\NormalTok{ pCensoring)}
\NormalTok{      group2 <-}\StringTok{ }\KeywordTok{interimCensoring}\NormalTok{(group0, mediana, Q99, }\DataTypeTok{pCensoring =}\NormalTok{ pCensoring)}
\NormalTok{      group3 <-}\StringTok{ }\KeywordTok{caseCensoring}\NormalTok{(group0, }\DataTypeTok{pCensoring =}\NormalTok{ pCensoring)}
      \KeywordTok{trialPlot}\NormalTok{(group0, group1, }\DataTypeTok{main =} \StringTok{"Time censoring"}\NormalTok{, }\DataTypeTok{col =} \StringTok{"red"}\NormalTok{, }\DataTypeTok{line =} \DecValTok{-1}\NormalTok{)}
      \KeywordTok{trialPlot}\NormalTok{(group0, group2, }\DataTypeTok{main =} \StringTok{"Interim censoring"}\NormalTok{, }\DataTypeTok{col =} \StringTok{"blue"}\NormalTok{, }\DataTypeTok{line =} \DecValTok{-1}\NormalTok{)}
      \KeywordTok{trialPlot}\NormalTok{(group0, group3, }\DataTypeTok{main =} \StringTok{"Case censoring"}\NormalTok{, }\DataTypeTok{col =} \StringTok{"green"}\NormalTok{, }\DataTypeTok{line =} \DecValTok{-1}\NormalTok{)}
\NormalTok{\}}
\end{Highlighting}
\end{Shaded}

\vspace{5mm}

This function plots a virtual experiment as a scatter plot of
\emph{hazard ratio} vs \emph{proportion of censoring} where every
experiment is a dot with the corresponding color to the type of
censoring (red for \emph{time censoring}, blue for \emph{interim
censoring}, and green for \emph{case censoring})

@param df The experiment dataframe. Every row, a virtual trial, is
represented as a dot

@param cols The colores to be used for the 3 types of censoring

@return a scatter plot

\vspace{2mm}

\begin{Shaded}
\begin{Highlighting}[]
\NormalTok{censorPlot <-}\StringTok{ }\ControlFlowTok{function}\NormalTok{(df, }\DataTypeTok{cols =} \KeywordTok{c}\NormalTok{(}\StringTok{"red"}\NormalTok{, }\StringTok{"blue"}\NormalTok{, }\StringTok{"green"}\NormalTok{), ...) \{}
\NormalTok{   types <-}\StringTok{  }\KeywordTok{c}\NormalTok{(}\StringTok{"time"}\NormalTok{, }\StringTok{"interim"}\NormalTok{, }\StringTok{"case"}\NormalTok{)}
\NormalTok{   pch <-}\StringTok{ }\KeywordTok{ifelse}\NormalTok{(df}\OperatorTok{$}\NormalTok{pValue }\OperatorTok{<}\StringTok{ }\FloatTok{0.05}\NormalTok{, }\DecValTok{19}\NormalTok{, }\DecValTok{21}\NormalTok{)}
\NormalTok{   col <-}\StringTok{ }\NormalTok{cols[}\KeywordTok{factor}\NormalTok{(df}\OperatorTok{$}\NormalTok{type, }\DataTypeTok{levels =}\NormalTok{ types)]}
   \KeywordTok{with}\NormalTok{(df, }\KeywordTok{plot}\NormalTok{(pCensored, hr, }\DataTypeTok{las =} \DecValTok{1}\NormalTok{, }\DataTypeTok{pch =}\NormalTok{ pch, }\DataTypeTok{col =}\NormalTok{ col, }\DataTypeTok{ylim =} \KeywordTok{c}\NormalTok{(}\DecValTok{0}\NormalTok{, }\FloatTok{1.2}\NormalTok{),  ...))}
\NormalTok{   caseCensored <-}\StringTok{ }\NormalTok{df[df[, }\StringTok{"type"}\NormalTok{] }\OperatorTok{==}\StringTok{ "case"}\NormalTok{, ]}
\NormalTok{   lm1 <-}\StringTok{ }\KeywordTok{lm}\NormalTok{(}\DataTypeTok{formula =}\NormalTok{ hr }\OperatorTok{~}\StringTok{ }\NormalTok{pCensored, }\DataTypeTok{data =}\NormalTok{ caseCensored)}
   \KeywordTok{abline}\NormalTok{(}\DataTypeTok{h =} \DecValTok{1}\NormalTok{, lm1)}
\NormalTok{   df}\OperatorTok{$}\NormalTok{sig <-}\StringTok{ }\NormalTok{(df}\OperatorTok{$}\NormalTok{pValue }\OperatorTok{<}\StringTok{ }\FloatTok{0.05}\NormalTok{) }\OperatorTok{*}\StringTok{ }\DecValTok{1}
\NormalTok{   leg <-}\StringTok{ }\KeywordTok{legend}\NormalTok{(}\StringTok{"bottomleft"}\NormalTok{, }\DataTypeTok{bty =} \StringTok{"n"}\NormalTok{, }\KeywordTok{paste}\NormalTok{(types, }\StringTok{"censoring"}\NormalTok{))}
   \KeywordTok{points}\NormalTok{(}\DataTypeTok{x =}\NormalTok{ leg}\OperatorTok{$}\NormalTok{text}\OperatorTok{$}\NormalTok{x }\OperatorTok{+}\StringTok{ }\NormalTok{leg}\OperatorTok{$}\NormalTok{rect}\OperatorTok{$}\NormalTok{w, }\DataTypeTok{y =}\NormalTok{ leg}\OperatorTok{$}\NormalTok{text}\OperatorTok{$}\NormalTok{y, }\DataTypeTok{col =}\NormalTok{ cols, }\DataTypeTok{pch =} \DecValTok{19}\NormalTok{)}
   \KeywordTok{points}\NormalTok{(}\DataTypeTok{x =}\NormalTok{ leg}\OperatorTok{$}\NormalTok{text}\OperatorTok{$}\NormalTok{x }\OperatorTok{+}\StringTok{ }\NormalTok{leg}\OperatorTok{$}\NormalTok{rect}\OperatorTok{$}\NormalTok{w }\OperatorTok{*}\StringTok{ }\FloatTok{1.3}\NormalTok{, }\DataTypeTok{y =}\NormalTok{ leg}\OperatorTok{$}\NormalTok{text}\OperatorTok{$}\NormalTok{y, }\DataTypeTok{col =}\NormalTok{ cols)}
   \KeywordTok{text}\NormalTok{  (}\DataTypeTok{x =}\NormalTok{ leg}\OperatorTok{$}\NormalTok{text}\OperatorTok{$}\NormalTok{x }\OperatorTok{-}\StringTok{ }\FloatTok{.05}\NormalTok{, }\DataTypeTok{y =}\NormalTok{ leg}\OperatorTok{$}\NormalTok{text}\OperatorTok{$}\NormalTok{y[}\DecValTok{1}\NormalTok{] }\OperatorTok{+}\StringTok{ }\FloatTok{0.05}\NormalTok{, }\StringTok{"p-value  <0.05  >0.05"}\NormalTok{,}
          \DataTypeTok{pos =} \DecValTok{4}\NormalTok{, }\DataTypeTok{offset =} \DecValTok{6}\NormalTok{, ...)}
   \KeywordTok{legend}\NormalTok{(}\StringTok{"bottomright"}\NormalTok{, }\DataTypeTok{bty =} \StringTok{"n"}\NormalTok{, }
          \DataTypeTok{legend =} \KeywordTok{paste}\NormalTok{(}\StringTok{"r ="}\NormalTok{, }\KeywordTok{with}\NormalTok{(caseCensored, }\KeywordTok{signif}\NormalTok{(}\KeywordTok{cor}\NormalTok{(hr, pCensored), }\DecValTok{3}\NormalTok{))))}
\NormalTok{\}}
\end{Highlighting}
\end{Shaded}

\vspace{5mm}

This function plots a virtual experiment as a scatter plot of
\emph{hazard ratio} vs \emph{bias index} where every experiment is a dot
with the corresponding color to the type of censoring (red for
\emph{time censoring}, blue for \emph{interim censoring}, and green for
\emph{case censoring})

@param df The experiment dataframe. Every row, a virtual trial, is
represented as a dot

@param cols The colores to be used for the 3 types of censoring

@return a plot A scatter plot of \emph{hazard ratio} vs \emph{proportion
of censoring}. Every virtual trial is a dot with the corresponding color
to the type of censoring (red for \emph{time censoring}, blue for
\emph{interim censoring}, and green for \emph{case censoring})

\vspace{2mm}

\begin{Shaded}
\begin{Highlighting}[]
\NormalTok{biasPlot <-}\StringTok{ }\ControlFlowTok{function}\NormalTok{(df, }\DataTypeTok{cols =} \KeywordTok{c}\NormalTok{(}\StringTok{"red"}\NormalTok{, }\StringTok{"blue"}\NormalTok{, }\StringTok{"green"}\NormalTok{), BI, ...) \{}
\NormalTok{   X <-}\StringTok{ }\NormalTok{df[, BI]}
\NormalTok{   types <-}\StringTok{  }\KeywordTok{c}\NormalTok{(}\StringTok{"time"}\NormalTok{, }\StringTok{"interim"}\NormalTok{, }\StringTok{"case"}\NormalTok{)}
\NormalTok{   pch <-}\StringTok{ }\KeywordTok{ifelse}\NormalTok{(df}\OperatorTok{$}\NormalTok{pValue }\OperatorTok{<}\StringTok{ }\FloatTok{0.05}\NormalTok{, }\DecValTok{19}\NormalTok{, }\DecValTok{1}\NormalTok{)}
\NormalTok{   col <-}\StringTok{ }\NormalTok{cols[}\KeywordTok{factor}\NormalTok{(df}\OperatorTok{$}\NormalTok{type, }\DataTypeTok{levels =}\NormalTok{ types)]}
   \KeywordTok{with}\NormalTok{(df, }\KeywordTok{plot}\NormalTok{(X, hr, }\DataTypeTok{las =} \DecValTok{1}\NormalTok{, }\DataTypeTok{pch =}\NormalTok{ pch, }\DataTypeTok{col =}\NormalTok{ col, }\DataTypeTok{xlab =}\NormalTok{ BI, }\DataTypeTok{ylim =} \KeywordTok{c}\NormalTok{(}\DecValTok{0}\NormalTok{, }\FloatTok{1.6}\NormalTok{), ...))}
\NormalTok{   leg1 <-}\StringTok{ }\KeywordTok{legend}\NormalTok{(}\StringTok{"topleft"}\NormalTok{, }\StringTok{"               p-value  <0.05  >0.05"}\NormalTok{, }\DataTypeTok{bty =} \StringTok{"n"}\NormalTok{)}
\NormalTok{   leg <-}\StringTok{ }\KeywordTok{legend}\NormalTok{(}\DataTypeTok{x =}\NormalTok{ leg1}\OperatorTok{$}\NormalTok{rect}\OperatorTok{$}\NormalTok{left, }\DataTypeTok{y =}\NormalTok{ leg1}\OperatorTok{$}\NormalTok{text}\OperatorTok{$}\NormalTok{y, }\DataTypeTok{bty =} \StringTok{"n"}\NormalTok{, }\KeywordTok{paste}\NormalTok{(types, }\StringTok{"censoring"}\NormalTok{))}
   \KeywordTok{points}\NormalTok{(}\DataTypeTok{x =}\NormalTok{ leg}\OperatorTok{$}\NormalTok{text}\OperatorTok{$}\NormalTok{x }\OperatorTok{+}\StringTok{ }\NormalTok{leg}\OperatorTok{$}\NormalTok{rect}\OperatorTok{$}\NormalTok{w, }\DataTypeTok{y =}\NormalTok{ leg}\OperatorTok{$}\NormalTok{text}\OperatorTok{$}\NormalTok{y, }\DataTypeTok{col =}\NormalTok{ cols, }\DataTypeTok{pch =} \DecValTok{19}\NormalTok{)}
   \KeywordTok{points}\NormalTok{(}\DataTypeTok{x =}\NormalTok{ leg}\OperatorTok{$}\NormalTok{text}\OperatorTok{$}\NormalTok{x }\OperatorTok{+}\StringTok{ }\NormalTok{leg}\OperatorTok{$}\NormalTok{rect}\OperatorTok{$}\NormalTok{w }\OperatorTok{*}\StringTok{ }\FloatTok{1.3}\NormalTok{, }\DataTypeTok{y =}\NormalTok{ leg}\OperatorTok{$}\NormalTok{text}\OperatorTok{$}\NormalTok{y, }\DataTypeTok{col =}\NormalTok{ cols)}
\NormalTok{   df}\OperatorTok{$}\NormalTok{sig <-}\StringTok{ }\NormalTok{(df}\OperatorTok{$}\NormalTok{pValue }\OperatorTok{<}\StringTok{ }\FloatTok{0.05}\NormalTok{) }\OperatorTok{*}\StringTok{ }\DecValTok{1}
\NormalTok{   oCutPoints <-}\StringTok{ }\KeywordTok{optimal.cutpoints}\NormalTok{(}
      \DataTypeTok{X =}\NormalTok{ BI, }\DataTypeTok{status =} \StringTok{"sig"}\NormalTok{, }\DataTypeTok{data =}\NormalTok{ df, }\DataTypeTok{tag.healthy =} \DecValTok{0}\NormalTok{, }\DataTypeTok{methods =} \StringTok{"Youden"}\NormalTok{)}
\NormalTok{   sCutPoints <-}\StringTok{ }\KeywordTok{summary}\NormalTok{(oCutPoints)}\OperatorTok{$}\NormalTok{p.table}\OperatorTok{$}\NormalTok{Global}
\NormalTok{   roc <-}\StringTok{ }\KeywordTok{legend}\NormalTok{(}\StringTok{"topright"}\NormalTok{, }\DataTypeTok{bty =} \StringTok{"n"}\NormalTok{, }\DataTypeTok{xjust =} \FloatTok{0.5}\NormalTok{,}
                 \DataTypeTok{legend =} \KeywordTok{c}\NormalTok{(}\StringTok{"Area under ROC"}\NormalTok{, sCutPoints}\OperatorTok{$}\NormalTok{AUC_CI))}
\NormalTok{   params <-}\StringTok{ }\KeywordTok{signif}\NormalTok{(sCutPoints}\OperatorTok{$}\NormalTok{Youden[[}\DecValTok{1}\NormalTok{]][}\DecValTok{1}\OperatorTok{:}\DecValTok{5}\NormalTok{, ], }\DecValTok{3}\NormalTok{)}
\NormalTok{   cutoff <-}\StringTok{ }\NormalTok{params[[}\DecValTok{1}\NormalTok{]]}
\NormalTok{   params <-}\StringTok{ }\KeywordTok{paste}\NormalTok{(}\KeywordTok{c}\NormalTok{(}\StringTok{"Cutoff"}\NormalTok{, }\StringTok{"Sensit"}\NormalTok{, }\StringTok{"Specif"}\NormalTok{, }\StringTok{"posPV"}\NormalTok{, }\StringTok{"negPV"}\NormalTok{), params, }\DataTypeTok{sep =} \StringTok{": "}\NormalTok{)}
   \KeywordTok{abline}\NormalTok{(}\DataTypeTok{h =} \DecValTok{1}\NormalTok{, }\DataTypeTok{v =}\NormalTok{ cutoff)}
\NormalTok{   threshold <-}\StringTok{ }\KeywordTok{min}\NormalTok{(BI[df}\OperatorTok{$}\NormalTok{pValue }\OperatorTok{<}\StringTok{ }\FloatTok{0.05}\NormalTok{], }\DataTypeTok{na.rm =} \OtherTok{FALSE}\NormalTok{)}
   \KeywordTok{legend}\NormalTok{(}\StringTok{"bottomleft"}\NormalTok{, }\DataTypeTok{legend =}\NormalTok{ params, }\DataTypeTok{bty =} \StringTok{"n"}\NormalTok{)}
\NormalTok{\}}
\end{Highlighting}
\end{Shaded}

\hypertarget{figures}{%
\subsection{Figures}\label{figures}}

\hypertarget{figure-1}{%
\subsubsection{Figure 1}\label{figure-1}}

Simulation of 12 virtual experiments based on 4 complete follow-up
datasets with median survival of 25 and 50 time units
(\(Q_{99} = 100\)), and cure-rate of 0 and 0.4. For every complete
follow-up dataset with 1000 cases, three censored datasets are obtained
with the three different mechanisms of censoring and proportion of
censoring of 50\%.

\vspace{2mm}

\begin{Shaded}
\begin{Highlighting}[]
\KeywordTok{set.seed}\NormalTok{(}\DecValTok{1963}\NormalTok{)}
\NormalTok{op <-}\StringTok{ }\KeywordTok{par}\NormalTok{(}\DataTypeTok{mfrow =} \KeywordTok{c}\NormalTok{(}\DecValTok{4}\NormalTok{, }\DecValTok{3}\NormalTok{), }\DataTypeTok{cex =} \FloatTok{0.8}\NormalTok{)}
   \KeywordTok{virtualTrials}\NormalTok{(}\DataTypeTok{pCure =} \DecValTok{0}\NormalTok{, }\DataTypeTok{mediana =} \DecValTok{25}\NormalTok{)}
   \KeywordTok{virtualTrials}\NormalTok{(}\DataTypeTok{pCure =} \DecValTok{0}\NormalTok{, }\DataTypeTok{mediana =} \DecValTok{50}\NormalTok{)}
   \KeywordTok{virtualTrials}\NormalTok{(}\DataTypeTok{pCure =} \FloatTok{0.4}\NormalTok{, }\DataTypeTok{mediana =} \DecValTok{50}\NormalTok{)}
   \KeywordTok{virtualTrials}\NormalTok{(}\DataTypeTok{pCure =} \FloatTok{0.4}\NormalTok{, }\DataTypeTok{mediana =} \DecValTok{50}\NormalTok{)}
\KeywordTok{par}\NormalTok{(op)}
\end{Highlighting}
\end{Shaded}

\hypertarget{virtual-experiments}{%
\subsubsection{Virtual experiments}\label{virtual-experiments}}

Simulates different experiments, every one with 3000 virtual trials
based on 1000 random complete follow-up datasets with a range of median
distribution of events and cure-rates

\begin{Shaded}
\begin{Highlighting}[]
\NormalTok{n =}\StringTok{ }\DecValTok{1000}
\KeywordTok{set.seed}\NormalTok{(}\DecValTok{1963}\NormalTok{); experiment1 <-}
\StringTok{   }\KeywordTok{simulTrials}\NormalTok{(}\DataTypeTok{nTrials =}\NormalTok{ n, }\DataTypeTok{mediana =} \StringTok{"runif(1, 5, 50)"}\NormalTok{, }\DataTypeTok{pCure =} \StringTok{"0"}\NormalTok{)}
\KeywordTok{set.seed}\NormalTok{(}\DecValTok{1963}\NormalTok{); experiment2 <-}
\StringTok{   }\KeywordTok{simulTrials}\NormalTok{(}\DataTypeTok{nTrials =}\NormalTok{ n, }\DataTypeTok{mediana =} \StringTok{"runif(1, 50, 95)"}\NormalTok{, }\DataTypeTok{pCure =} \StringTok{"0"}\NormalTok{)}
\KeywordTok{set.seed}\NormalTok{(}\DecValTok{1963}\NormalTok{); experiment3 <-}
\StringTok{   }\KeywordTok{simulTrials}\NormalTok{(}\DataTypeTok{nTrials =}\NormalTok{ n, }\DataTypeTok{mediana =} \StringTok{"runif(1, 5, 50)"}\NormalTok{, }\DataTypeTok{pCure =} \StringTok{"0.5"}\NormalTok{)}
\KeywordTok{set.seed}\NormalTok{(}\DecValTok{1963}\NormalTok{); experiment4 <-}
\StringTok{   }\KeywordTok{simulTrials}\NormalTok{(}\DataTypeTok{nTrials =}\NormalTok{ n, }\DataTypeTok{mediana =} \StringTok{"runif(1, 50, 95)"}\NormalTok{, }\DataTypeTok{pCure =} \StringTok{"0.5"}\NormalTok{)}
\KeywordTok{set.seed}\NormalTok{(}\DecValTok{1963}\NormalTok{); experiment5 <-}
\StringTok{   }\KeywordTok{simulTrials}\NormalTok{(}\DataTypeTok{nTrials =}\NormalTok{ n, }\DataTypeTok{mediana =} \StringTok{"runif(1, 5, 95)"}\NormalTok{, }\DataTypeTok{pCure =} \StringTok{"runif(1, 0, 0.8)"}\NormalTok{)}
\end{Highlighting}
\end{Shaded}

\hypertarget{figure-2}{%
\subsubsection{Figure 2}\label{figure-2}}

Simulation of 3000 virtual trials based on 1000 random complete
follow-up datasets with medians of 5 to 95 time units and cure-rate of 0
to 0.8; time censoring (red), interim censoring (blue) and case
censoring (green) datasets are simulated with probability of censoring
of 5\% to 95\%.

\vspace{2mm}

\begin{Shaded}
\begin{Highlighting}[]
\NormalTok{op <-}\StringTok{ }\KeywordTok{par}\NormalTok{(}\DataTypeTok{cex =} \FloatTok{0.8}\NormalTok{)}
   \KeywordTok{censorPlot}\NormalTok{(experiment5, }\DataTypeTok{xlim =} \KeywordTok{c}\NormalTok{(}\FloatTok{0.075}\NormalTok{, }\FloatTok{0.925}\NormalTok{), }\DataTypeTok{xlab =} \StringTok{"proportion of censoring"}\NormalTok{,}
              \DataTypeTok{main =} \StringTok{"median: 5-95   pCure = 0.-0.8"}\NormalTok{)}
\KeywordTok{par}\NormalTok{(op)}
\end{Highlighting}
\end{Shaded}

\hypertarget{figure-3}{%
\subsubsection{Figure 3}\label{figure-3}}

Simulation of 2 experiments, each of them with 3000 trials based on 1000
complete follow-up datasets and the 3 mechanisms of censoring described,
with random proportion of censoring in the range 0.05 to 0.95.

\vspace{2mm}

\begin{Shaded}
\begin{Highlighting}[]
\NormalTok{op <-}\StringTok{ }\KeywordTok{par}\NormalTok{(}\DataTypeTok{mfrow =} \KeywordTok{c}\NormalTok{(}\DecValTok{1}\NormalTok{, }\DecValTok{2}\NormalTok{))}
   \KeywordTok{biasPlot}\NormalTok{(experiment1, }\DataTypeTok{BI =} \StringTok{"SQBI"}\NormalTok{,  }\DataTypeTok{main =} \StringTok{"median: 5-50   pCure = 0"}\NormalTok{, }\DataTypeTok{xlim =} \KeywordTok{c}\NormalTok{(}\FloatTok{0.2}\NormalTok{, }\FloatTok{1.9}\NormalTok{))}
   \KeywordTok{biasPlot}\NormalTok{(experiment5, }\DataTypeTok{BI =} \StringTok{"SQBI"}\NormalTok{,  }\DataTypeTok{main =} \StringTok{"median: 5-95   pCure = 0-0.8"}\NormalTok{)}
\KeywordTok{par}\NormalTok{(op)}
\end{Highlighting}
\end{Shaded}

\hypertarget{figure-4}{%
\subsubsection{Figure 4}\label{figure-4}}

Simulation of 4 virtual experiments with a 2x2 design: right skewed
Weibull distributions or not, and cure-rate of 0 or 50\%.

\vspace{2mm}

\begin{Shaded}
\begin{Highlighting}[]
\NormalTok{op <-}\StringTok{ }\KeywordTok{par}\NormalTok{(}\DataTypeTok{mfrow =} \KeywordTok{c}\NormalTok{(}\DecValTok{2}\NormalTok{, }\DecValTok{2}\NormalTok{))}
   \KeywordTok{biasPlot}\NormalTok{(experiment1, }\DataTypeTok{BI =} \StringTok{"UMBI"}\NormalTok{, }\DataTypeTok{main =} \StringTok{"median:  5-50   pCure = 0."}\NormalTok{)}
   \KeywordTok{biasPlot}\NormalTok{(experiment2, }\DataTypeTok{BI =} \StringTok{"UMBI"}\NormalTok{, }\DataTypeTok{main =} \StringTok{"median: 50-95   pCure = 0."}\NormalTok{)}
   \KeywordTok{biasPlot}\NormalTok{(experiment3, }\DataTypeTok{BI =} \StringTok{"UMBI"}\NormalTok{, }\DataTypeTok{main =} \StringTok{"median:  5-50   pCure = 0.5"}\NormalTok{)}
   \KeywordTok{biasPlot}\NormalTok{(experiment4, }\DataTypeTok{BI =} \StringTok{"UMBI"}\NormalTok{, }\DataTypeTok{main =} \StringTok{"median: 50-95   pCure = 0.5"}\NormalTok{)}
\KeywordTok{par}\NormalTok{(op)}
\end{Highlighting}
\end{Shaded}

\hypertarget{figure-5}{%
\subsubsection{Figure 5}\label{figure-5}}

Simulation of 2 experiments, each of them with 3000 trials based on 1000
complete follow-up datasets and the 3 mechanisms of censoring described,
with random proportion of censoring in the range 0.05 to 0.95.

\vspace{2mm}

\begin{Shaded}
\begin{Highlighting}[]
\NormalTok{op <-}\StringTok{ }\KeywordTok{par}\NormalTok{(}\DataTypeTok{mfrow =} \KeywordTok{c}\NormalTok{(}\DecValTok{1}\NormalTok{, }\DecValTok{2}\NormalTok{))}
   \KeywordTok{biasPlot}\NormalTok{(experiment1, }\DataTypeTok{BI =} \StringTok{"SABI"}\NormalTok{,  }\DataTypeTok{main =} \StringTok{"median: 5-50   pCure = 0"}\NormalTok{, }\DataTypeTok{xlim =} \KeywordTok{c}\NormalTok{(}\FloatTok{0.2}\NormalTok{, }\FloatTok{1.9}\NormalTok{))}
   \KeywordTok{biasPlot}\NormalTok{(experiment5, }\DataTypeTok{BI =} \StringTok{"SABI"}\NormalTok{,  }\DataTypeTok{main =} \StringTok{"median: 5-95   pCure = 0-0.8"}\NormalTok{, }\DataTypeTok{xlim =} \KeywordTok{c}\NormalTok{(}\FloatTok{0.2}\NormalTok{, }\FloatTok{2.5}\NormalTok{))}
\KeywordTok{par}\NormalTok{(op)}
\end{Highlighting}
\end{Shaded}

\hypertarget{clinical-available-datasets}{%
\subsection{Clinical available
datasets}\label{clinical-available-datasets}}

Available clinical datasets were retrieved from Documentation for
package \(survival\) version 2.44-1.1, based on 6 published articles
(\url{https://stat.ethz.ch/R-manual/R-devel/library/survival/html/00Index.html}).
Bias indexes SQBI and SABI were applied to the datasets and to the
research arms where available.

\hypertarget{acute-myelogenous-leukemia-survival-data}{%
\subsubsection{Acute Myelogenous Leukemia survival
data}\label{acute-myelogenous-leukemia-survival-data}}

\url{https://stat.ethz.ch/R-manual/R-devel/library/survival/html/aml.html}

\vspace{2mm}

\begin{Shaded}
\begin{Highlighting}[]
\NormalTok{trialsBI <-}\StringTok{ }\KeywordTok{clinicalBias}\NormalTok{(aml, }\StringTok{"Acute Myelogenous Leukemia survival data"}\NormalTok{, }\StringTok{"Miller1997"}\NormalTok{)}
\end{Highlighting}
\end{Shaded}

\hypertarget{bladder-cancer-recurrences}{%
\subsubsection{Bladder Cancer
Recurrences}\label{bladder-cancer-recurrences}}

\url{https://stat.ethz.ch/R-manual/R-devel/library/survival/html/bladder.html}

\vspace{2mm}

\begin{Shaded}
\begin{Highlighting}[]
\KeywordTok{data}\NormalTok{(bladder)}
\NormalTok{bladder1}\OperatorTok{$}\NormalTok{time   <-}\StringTok{ }\KeywordTok{with}\NormalTok{(bladder1, stop }\OperatorTok{-}\StringTok{ }\NormalTok{start)}
\NormalTok{bladder1}\OperatorTok{$}\NormalTok{status <-}\StringTok{ }\NormalTok{(bladder1}\OperatorTok{$}\NormalTok{status }\OperatorTok{==}\StringTok{ }\DecValTok{1} \OperatorTok{|}\StringTok{ }\NormalTok{bladder1}\OperatorTok{$}\NormalTok{status }\OperatorTok{==}\StringTok{ }\DecValTok{2}\NormalTok{) }\OperatorTok{*}\StringTok{ }\DecValTok{1}
\KeywordTok{names}\NormalTok{(bladder1)[}\DecValTok{2}\NormalTok{] <-}\StringTok{ "x"}
\NormalTok{trialsBI <-}\StringTok{ }\KeywordTok{rbind}\NormalTok{(trialsBI,}
                  \KeywordTok{clinicalBias}\NormalTok{(bladder1, }\StringTok{"Bladder Cancer Recurrences - bladder1"}\NormalTok{, }\StringTok{"Wei1989"}\NormalTok{))}
\NormalTok{placebo    <-}\StringTok{ }\NormalTok{bladder1[bladder1}\OperatorTok{$}\NormalTok{x }\OperatorTok{==}\StringTok{ "placebo"}\NormalTok{, ]}
\NormalTok{pyridoxine <-}\StringTok{ }\NormalTok{bladder1[bladder1}\OperatorTok{$}\NormalTok{x }\OperatorTok{==}\StringTok{ "pyridoxine"}\NormalTok{, ]}
\NormalTok{thiotepa   <-}\StringTok{ }\NormalTok{bladder1[bladder1}\OperatorTok{$}\NormalTok{x }\OperatorTok{==}\StringTok{ "thiotepa"}\NormalTok{, ]}
\NormalTok{trialsBI <-}\StringTok{ }\KeywordTok{rbind}\NormalTok{(trialsBI, }\KeywordTok{clinicalBias}\NormalTok{(placebo,    }\StringTok{" - placebo"}\NormalTok{, }\StringTok{""}\NormalTok{))}
\NormalTok{trialsBI <-}\StringTok{ }\KeywordTok{rbind}\NormalTok{(trialsBI, }\KeywordTok{clinicalBias}\NormalTok{(pyridoxine, }\StringTok{" - pyridoxine"}\NormalTok{, }\StringTok{""}\NormalTok{))}
\NormalTok{trialsBI <-}\StringTok{ }\KeywordTok{rbind}\NormalTok{(trialsBI, }\KeywordTok{clinicalBias}\NormalTok{(thiotepa,   }\StringTok{" - thiotepa"}\NormalTok{, }\StringTok{""}\NormalTok{))}

\NormalTok{bladder2}\OperatorTok{$}\NormalTok{time <-}\StringTok{ }\KeywordTok{with}\NormalTok{(bladder2, stop }\OperatorTok{-}\StringTok{ }\NormalTok{start)}
\KeywordTok{names}\NormalTok{(bladder2)[}\DecValTok{2}\NormalTok{] <-}\StringTok{ "x"}
\KeywordTok{names}\NormalTok{(bladder2)[}\DecValTok{7}\NormalTok{] <-}\StringTok{ "status"}
\NormalTok{trialsBI <-}\StringTok{ }\KeywordTok{rbind}\NormalTok{(trialsBI,}
                  \KeywordTok{clinicalBias}\NormalTok{(bladder2, }\StringTok{"Bladder Cancer Recurrences - bladder2"}\NormalTok{, }\StringTok{"Wei1989"}\NormalTok{))}
\NormalTok{rx1 <-}\StringTok{ }\NormalTok{bladder2[bladder2}\OperatorTok{$}\NormalTok{x }\OperatorTok{==}\StringTok{ }\DecValTok{1}\NormalTok{, ]}
\NormalTok{rx2 <-}\StringTok{ }\NormalTok{bladder2[bladder2}\OperatorTok{$}\NormalTok{x }\OperatorTok{==}\StringTok{ }\DecValTok{2}\NormalTok{, ]}
\NormalTok{trialsBI <-}\StringTok{ }\KeywordTok{rbind}\NormalTok{(trialsBI, }\KeywordTok{clinicalBias}\NormalTok{(rx1, }\StringTok{" - rx1"}\NormalTok{, }\StringTok{""}\NormalTok{))}
\NormalTok{trialsBI <-}\StringTok{ }\KeywordTok{rbind}\NormalTok{(trialsBI, }\KeywordTok{clinicalBias}\NormalTok{(rx2, }\StringTok{" - rx2"}\NormalTok{, }\StringTok{""}\NormalTok{))}
\end{Highlighting}
\end{Shaded}

\hypertarget{ncctg-lung-cancer-data}{%
\subsubsection{NCCTG Lung Cancer Data}\label{ncctg-lung-cancer-data}}

\url{https://stat.ethz.ch/R-manual/R-devel/library/survival/html/lung.html}

\vspace{2mm}

\begin{Shaded}
\begin{Highlighting}[]
\KeywordTok{data}\NormalTok{(lung)}
\NormalTok{lung}\OperatorTok{$}\NormalTok{status <-}\StringTok{ }\NormalTok{(lung}\OperatorTok{$}\NormalTok{status }\OperatorTok{==}\StringTok{ }\DecValTok{2}\NormalTok{) }\OperatorTok{*}\StringTok{ }\DecValTok{1}
\NormalTok{lung}\OperatorTok{$}\NormalTok{x =}\StringTok{ ""}
\NormalTok{trialsBI <-}\StringTok{ }\KeywordTok{rbind}\NormalTok{(trialsBI, }\KeywordTok{clinicalBias}\NormalTok{(lung, }\StringTok{"NCCTG Lung Cancer Data"}\NormalTok{, }\StringTok{"Loprinzi1994"}\NormalTok{))}
\end{Highlighting}
\end{Shaded}

\hypertarget{chemotherapy-for-stage-bc-colon-cancer}{%
\subsubsection{Chemotherapy for Stage B/C colon
cancer}\label{chemotherapy-for-stage-bc-colon-cancer}}

\url{https://stat.ethz.ch/R-manual/R-devel/library/survival/html/colon.html}

\vspace{2mm}

\begin{Shaded}
\begin{Highlighting}[]
\KeywordTok{data}\NormalTok{(colon)}
\NormalTok{colon}\OperatorTok{$}\NormalTok{x <-}\StringTok{ }\NormalTok{colon}\OperatorTok{$}\NormalTok{rx}
\NormalTok{trialsBI <-}\StringTok{ }\KeywordTok{rbind}\NormalTok{(trialsBI,}
                  \KeywordTok{clinicalBias}\NormalTok{(colon, }\StringTok{"Chemotherapy for Stage B/C colon cancer"}\NormalTok{, }\StringTok{"Moertel1990"}\NormalTok{))}
\NormalTok{Obs <-}\StringTok{ }\NormalTok{colon[colon}\OperatorTok{$}\NormalTok{rx }\OperatorTok{==}\StringTok{ "Obs"}\NormalTok{, ]}
\NormalTok{Lev <-}\StringTok{ }\NormalTok{colon[colon}\OperatorTok{$}\NormalTok{rx }\OperatorTok{==}\StringTok{ "Lev"}\NormalTok{, ]}
\NormalTok{LFU <-}\StringTok{ }\NormalTok{colon[colon}\OperatorTok{$}\NormalTok{rx }\OperatorTok{==}\StringTok{ "Lev+5FU"}\NormalTok{, ]}
\NormalTok{trialsBI <-}\StringTok{ }\KeywordTok{rbind}\NormalTok{(trialsBI, }\KeywordTok{clinicalBias}\NormalTok{(Obs, }\StringTok{" - Observation"}\NormalTok{, }\StringTok{""}\NormalTok{))}
\NormalTok{trialsBI <-}\StringTok{ }\KeywordTok{rbind}\NormalTok{(trialsBI, }\KeywordTok{clinicalBias}\NormalTok{(Lev, }\StringTok{" - Levamisol"}\NormalTok{, }\StringTok{""}\NormalTok{))}
\NormalTok{trialsBI <-}\StringTok{ }\KeywordTok{rbind}\NormalTok{(trialsBI, }\KeywordTok{clinicalBias}\NormalTok{(LFU, }\StringTok{" - Levamisol + 5FU"}\NormalTok{, }\StringTok{""}\NormalTok{))}
\end{Highlighting}
\end{Shaded}

\hypertarget{ovarian-cancer-survival-data}{%
\subsubsection{Ovarian Cancer Survival
Data}\label{ovarian-cancer-survival-data}}

\url{https://stat.ethz.ch/R-manual/R-devel/library/survival/html/ovarian.html}

\vspace{2mm}

\begin{Shaded}
\begin{Highlighting}[]
\KeywordTok{data}\NormalTok{(ovarian)}
\NormalTok{ovarian}\OperatorTok{$}\NormalTok{time   <-}\StringTok{ }\NormalTok{ovarian}\OperatorTok{$}\NormalTok{futime}
\NormalTok{ovarian}\OperatorTok{$}\NormalTok{status <-}\StringTok{ }\NormalTok{ovarian}\OperatorTok{$}\NormalTok{fustat}
\NormalTok{ovarian}\OperatorTok{$}\NormalTok{x      <-}\StringTok{ }\NormalTok{ovarian}\OperatorTok{$}\NormalTok{rx}
\NormalTok{trialsBI <-}\StringTok{ }\KeywordTok{rbind}\NormalTok{(trialsBI,}
                  \KeywordTok{clinicalBias}\NormalTok{(ovarian, }\StringTok{"Ovarian Cancer Survival Data"}\NormalTok{, }\StringTok{"Edmonson1979"}\NormalTok{))}
\NormalTok{rx1 <-}\StringTok{ }\NormalTok{ovarian[ovarian}\OperatorTok{$}\NormalTok{rx }\OperatorTok{==}\StringTok{ }\DecValTok{1}\NormalTok{, ]}
\NormalTok{rx2 <-}\StringTok{ }\NormalTok{ovarian[ovarian}\OperatorTok{$}\NormalTok{rx }\OperatorTok{==}\StringTok{ }\DecValTok{2}\NormalTok{, ]}
\NormalTok{trialsBI <-}\StringTok{ }\KeywordTok{rbind}\NormalTok{(trialsBI, }\KeywordTok{clinicalBias}\NormalTok{(rx1, }\StringTok{" - rx1"}\NormalTok{, }\StringTok{""}\NormalTok{))}
\NormalTok{trialsBI <-}\StringTok{ }\KeywordTok{rbind}\NormalTok{(trialsBI, }\KeywordTok{clinicalBias}\NormalTok{(rx2, }\StringTok{" - rx2"}\NormalTok{, }\StringTok{""}\NormalTok{))}
\end{Highlighting}
\end{Shaded}

\hypertarget{veterans-administration-lung-cancer-study}{%
\subsubsection{Veterans' Administration Lung Cancer
Study}\label{veterans-administration-lung-cancer-study}}

\url{https://stat.ethz.ch/R-manual/R-devel/library/survival/html/veteran.html}

\vspace{2mm}

\begin{Shaded}
\begin{Highlighting}[]
\KeywordTok{data}\NormalTok{(veteran)}
\NormalTok{veteran}\OperatorTok{$}\NormalTok{x <-}\StringTok{ }\NormalTok{veteran}\OperatorTok{$}\NormalTok{trt}
\NormalTok{trialsBI <-}\StringTok{ }\KeywordTok{rbind}\NormalTok{(trialsBI,}
            \KeywordTok{clinicalBias}\NormalTok{(veteran, }\StringTok{"Veterans' Administration Lung Cancer Study"}\NormalTok{, }\StringTok{"Kalbfleisch1980"}\NormalTok{))}
\NormalTok{standard <-}\StringTok{ }\NormalTok{veteran[veteran}\OperatorTok{$}\NormalTok{trt }\OperatorTok{==}\StringTok{ }\DecValTok{1}\NormalTok{, ]}
\NormalTok{test     <-}\StringTok{ }\NormalTok{veteran[veteran}\OperatorTok{$}\NormalTok{trt }\OperatorTok{==}\StringTok{ }\DecValTok{2}\NormalTok{, ]}
\NormalTok{trialsBI <-}\StringTok{ }\KeywordTok{rbind}\NormalTok{(trialsBI, }\KeywordTok{clinicalBias}\NormalTok{(standard, }\StringTok{" - standard"}\NormalTok{, }\StringTok{""}\NormalTok{))}
\NormalTok{trialsBI <-}\StringTok{ }\KeywordTok{rbind}\NormalTok{(trialsBI, }\KeywordTok{clinicalBias}\NormalTok{(test,     }\StringTok{" - test"}\NormalTok{, }\StringTok{""}\NormalTok{))}
\end{Highlighting}
\end{Shaded}

\hypertarget{table-1}{%
\subsubsection{Table 1}\label{table-1}}

\begin{Shaded}
\begin{Highlighting}[]
\NormalTok{table1 <-}\StringTok{ }\NormalTok{trialsBI}
\KeywordTok{rownames}\NormalTok{(table1) <-}\StringTok{ }\OtherTok{NULL}
\NormalTok{table1[, }\KeywordTok{c}\NormalTok{(}\StringTok{"pCens"}\NormalTok{, }\StringTok{"SQBI"}\NormalTok{, }\StringTok{"SABI"}\NormalTok{)] <-}\StringTok{ }\KeywordTok{signif}\NormalTok{(trialsBI[, }\KeywordTok{c}\NormalTok{(}\StringTok{"pCens"}\NormalTok{, }\StringTok{"SQBI"}\NormalTok{, }\StringTok{"SABI"}\NormalTok{)], }\DecValTok{2}\NormalTok{)}

\NormalTok{highlight <-}\StringTok{ }\ControlFlowTok{function}\NormalTok{(values, }\DataTypeTok{threshold =} \DecValTok{1}\NormalTok{) \{}
\NormalTok{   selected <-}\StringTok{ }\NormalTok{values }\OperatorTok{>}\StringTok{ }\NormalTok{threshold}
\NormalTok{   values[selected] <-}\StringTok{ }\KeywordTok{paste0}\NormalTok{(}\StringTok{"}\CharTok{\textbackslash{}\textbackslash{}}\StringTok{textbf\{"}\NormalTok{, values[selected], }\StringTok{"\}"}\NormalTok{)}
\NormalTok{   values}
\NormalTok{\}}
\NormalTok{table1[, }\StringTok{"SQBI"}\NormalTok{] <-}\StringTok{ }\KeywordTok{highlight}\NormalTok{(table1[, }\StringTok{"SQBI"}\NormalTok{], }\DecValTok{1}\NormalTok{)}
\NormalTok{table1[, }\StringTok{"SABI"}\NormalTok{] <-}\StringTok{ }\KeywordTok{highlight}\NormalTok{(table1[, }\StringTok{"SABI"}\NormalTok{], }\DecValTok{1}\NormalTok{)}

\KeywordTok{options}\NormalTok{(}\DataTypeTok{xtable.comment =} \OtherTok{FALSE}\NormalTok{)}
\NormalTok{xt <-}\StringTok{ }\KeywordTok{xtable}\NormalTok{(table1,  }\DataTypeTok{format =} \StringTok{"latex"}\NormalTok{, }\DataTypeTok{digits =} \DecValTok{2}\NormalTok{,}
             \DataTypeTok{caption =} \StringTok{"Bias in available clinical datasests"}\NormalTok{)}
\KeywordTok{print}\NormalTok{(xt, }\DataTypeTok{caption.placement =} \StringTok{"top"}\NormalTok{, }\DataTypeTok{type =} \StringTok{"latex"}\NormalTok{, }\DataTypeTok{include.rownames =} \OtherTok{FALSE}\NormalTok{,}
      \DataTypeTok{sanitize.text.function =}\NormalTok{ identity)}
\end{Highlighting}
\end{Shaded}

\hypertarget{sessioninfo}{%
\subsubsection{sessionInfo()}\label{sessioninfo}}

\begin{verbatim}
## R version 4.0.3 (2020-10-10)
## Platform: x86_64-apple-darwin17.0 (64-bit)
## Running under: macOS Catalina 10.15.7
## 
## Matrix products: default
## BLAS:   /Library/Frameworks/R.framework/Versions/4.0/Resources/lib/libRblas.dylib
## LAPACK: /Library/Frameworks/R.framework/Versions/4.0/Resources/lib/libRlapack.dylib
## 
## locale:
## [1] en_US.UTF-8/en_US.UTF-8/en_US.UTF-8/C/en_US.UTF-8/en_US.UTF-8
## 
## attached base packages:
## [1] stats     graphics  grDevices utils     datasets  methods   base     
## 
## other attached packages:
## [1] xtable_1.8-4           OptimalCutpoints_1.1-4 survival_3.2-7        
## [4] knitr_1.30            
## 
## loaded via a namespace (and not attached):
##  [1] lattice_0.20-41 digest_0.6.27   grid_4.0.3      magrittr_2.0.1 
##  [5] evaluate_0.14   rlang_0.4.9     stringi_1.5.3   Matrix_1.2-18  
##  [9] rmarkdown_2.5   splines_4.0.3   tools_4.0.3     stringr_1.4.0  
## [13] xfun_0.19       yaml_2.2.1      compiler_4.0.3  htmltools_0.5.0
\end{verbatim}